\documentclass[preprint]{aastex}

\newcommand{\kms}{\mbox{${\rm\,km\,s}^{-1}$}}

\newcommand{\kmspc}{\mbox{$\rm\,km\,s^{-1} pc^{-1}$}}

\newcommand{\kpc}{\mbox{$\rm\,kpc$}}
\newcommand{\msun}{\mbox{$\,M_\odot$}}

\newcommand{\sech}{\mbox{$\rm \,sech\,$}}

\newcommand{\magarc}{\mbox{${\rm\,mag\,arcsec}^{-2}$}}
\def\farcs{\hbox{$.\!\!^{\prime\prime}$}}
\def\arcdeg{\hbox{$^\circ$}}
\def\arcsec{\hbox{$^{\prime\prime}$}}

\begin{document}

\title{Self-Gravitating Eccentric Disk Models for the Double Nucleus of M31}

\author{Robert M. Salow\altaffilmark{1} and Thomas S.
Statler\altaffilmark{2}} \affil{Department of Physics and Astronomy,
Ohio University,  Athens, OH 45701}

\altaffiltext{1}{salow@helios.phy.ohiou.edu}
\altaffiltext{2}{statler@ohio.edu}

\slugcomment{Submitted to The Astrophysical Journal 2003 November 25}

\begin{abstract}
We present new dynamical models of weakly self-gravitating, finite
dispersion eccentric stellar disks around central black holes for the
double nucleus of M31.  The disk is fixed in a frame rotating at constant
precession speed, and is populated by stars on quasi-periodic orbits whose
parents are numerically integrated periodic orbits in the total potential. 
A distribution of quasi-periodic orbits about a given parent is
approximated by a distribution of Kepler orbits dispersed in eccentricity
and orientation, using an approximate phase-space distribution function
written in terms of the integrals of motion in the Kepler problem.  We use
these models, along with an optimization routine, to fit available
published kinematics and photometry in the inner $2\arcsec$ of the
nucleus.  A grid of 24 best-fit models is computed to accurately constrain
the mass of the central black hole and nuclear disk parameters.  We find
that the supermassive black hole in M31 has mass $M_{BH} = 5.62 \pm 0.66
\times 10^7 \msun$, which is consistent with the observed correlation
between the central black hole mass and the velocity dispersion of its
host spheroid.  Our models precess rapidly, at $\Omega = 36.5 \pm 4.2
\kmspc$, and possess a characteristic radial eccentricity distribution,
which gives rise to multi-modal line of sight velocity distributions along
lines of sight near the black hole.  These features can be used as
sensitive discriminants of disk structure. 
\end{abstract} 



\section{Introduction}
\label{sec:CHAP1}

It is widely believed that most, if not all, galaxies have supermassive
black holes (BHs) in their centers.  Estimates of the total mass density
in quasar remnants (Soltan 1982, Chokshi \& Turner 1992), models for the
evolution of the quasar luminosity function in hierarchical structure 
formation scenarios  (Haehnelt \& Rees 1993), and the large numbers of
nearby galaxies with low-luminosity nuclear activity (Ho, Filippenko \&
Sargent 1997) are consistent with this belief, assuming that active
galactic nuclei (AGNs) are powered by accretion of matter onto a BH
(Lynden-Bell 1969, Rees 1984). 

The discovery that the BH mass ($M_{BH}$) correlates with certain host
galaxy properties has made obtaining accurate masses for these objects a
high priority in extragalactic studies.  Kormendy \& Richstone (1995) and
Magorrian et al. (1998) find that BH mass is proportional to the mass or
luminosity of the host spheroidal component, though with significant
scatter.  Ferrarese \& Merritt (2000) and Gebhardt et al. (2000a) find that
the BH mass correlates with the velocity dispersion ($\sigma$) of the
stellar component, with much less scatter than the previous correlation;
from a sample of 31 galaxies with secure BH mass estimates, Tremaine et
al. (2002) find that the correlation can be written as $\log(M_{BH}/\msun)
= \alpha + \beta \log(\sigma/\sigma_0)$, where $\alpha = 8.13 \pm 0.06$
and $\beta = 4.02 \pm 0.32$ for a reference dispersion of $\sigma_0 = 200
\kms$.  Since $\sigma$ is measured outside the radius of influence
of the BH, defined to be $r_{h} = G M_{BH}/\sigma^2$, the $M_{BH}$ -
$\sigma$ correlation demonstrates a fundamental relationship between the
BH and its host spheroid.  Such a correlation has important implications
for theories of BH and galaxy formation and evolution.  It is thus
important to confirm and strengthen the correlation by providing
highly accurate BH masses for a large number of galaxies. 

BH masses can be found using a variety of techniques, including: 
measuring the kinematics of individually resolved stars (Eckart \& Genzel
1996, Ghez et al. 1998, Sch\"{o}del et al. 2002, Ghez et al. 2003),
dynamical modeling of spatially resolved stellar absorption-line
kinematics near the BH (see Kormendy \& Richstone 1995, Verolme et al.
2002, Gebhardt et al. 2003), measuring rotation curves from optical (Harms
et al. 1994, Macchetto et al.  1997, Bower et al. 1998, van der Marel \&
van den Bosch 1998, Marconi et al. 2003) or maser (Miyoshi et al. 1995,
Ishihara et al. 2001) emission lines from orbiting gas, reverberation
mapping in active galaxies (Peterson \& Wandel 1999, Peterson \& Wandel 2000,
Gebhardt et al. 2000b), and modeling of line profile widths (Vestergaard
2002).  The kinematics of resolved stellar motions and small maser disks
provide the most reliable mass estimates.  However, the motions of
individual stars can only be resolved in the Milky Way (see Sch\"{o}del et
al. 2002), and regular maser emission is only found in a few galaxies
(Hagiwara et al. 2003).  Of the other techniques, stellar-dynamical
modeling provides the most secure BH measurements;  gas near the BH can be
subject to non-gravitational forces, unlike the stellar component. 

The kinematic data must be resolved inside the region where Keplerian
motion dominates, however, to ensure that those stars fully contribute to
the line of sight velocity distribution (LOSVD), and not just in its
tails.  Simple calculations suggest that Keplerian motion should dominate
within a region of radius $r_k = k r_h$, where $k\approx 0.1$ - $0.3$,
depending on the BH mass and stellar radial density profile.  Using $k =
0.3$, it is easy to show that $r_k \leq 1.3 \times 10^{-3} M_{BH} /
\sigma^2$ (pc), with $M_{BH}$ in solar masses and $\sigma$ in km/s.  Using
values for $M_{BH}$, $\sigma$, and distance given in Tremaine et al.
(2002), or from the $M_{BH}$ - $\sigma$ relation, $r_k$ subtends an angle
of $11 \farcs 60$, $0\farcs 74$, and $0\farcs 15$ for the Milky Way, M31,
and M32, respectively.  Other than the Milky Way, M31 is the only nearby
galaxy in the Local Group with a resolved $r_k$ at the resolution of the
Hubble Space Telescope (HST) Space Telescope Imaging Spectrograph (STIS;
$\sim 0\farcs 1$); M33 is consistent with having a maximum BH mass of
$\sim 3000 \msun$ (Merritt et al. 2001, Gebhardt et al. 2001), so its
$r_k$ subtends an angle $< 0\farcs005$. 

M31 offers a unique opportunity to obtain a secure BH mass from spatially
resolved stellar kinematics inside $r_k$.  M31 is also the nearest galaxy
with a normal bulge (Kormendy 1993), and it has a nucleus; that is, a
small-scale stellar component which is photometrically and dynamically
distinct from the bulge and the large-scale galactic disk (Kormendy \&
Richstone 1995, and references therein).  Galaxy nuclei are poorly
understood, as is the dynamical connection between the nuclear stars and
the central BH.  M31's nucleus is $\sim 2\arcsec$ in radius (Light,
Danielson \& Schwarzschild 1974), which is fully within the sphere of
influence of the BH;  $r_h \simeq 2\farcs5$, if M31 is located at a
distance of $770$ kpc and has a BH mass of $5.5 \times 10^7 \msun$, as
implied by the $M_{BH}$ - $\sigma$ relation.  Thus, M31's nucleus allows
for a more detailed dynamical study than is possible for any other galaxy. 
Even more enticing is the fact that the nucleus is shown to be double; 
the photometric profile shows two brightness peaks, one of which is
off-center with respect to the outer bulge isophotes.  Kinematic profiles
also show strong asymmetries.  Thus, standard axisymmetric dynamical
modeling techniques (e.g. in Gebhardt et al. 2003) are inappropriate.  New
modeling methods are needed to obtain an accurate measure the BH mass in
M31. 

M31 was first shown to have a photometrically asymmetric nucleus by Light
et al. (1974), using the Stratoscope II balloon-borne telescope.  Nieto et
al. (1986) confirmed those observations with groundbased data, and found
that the brightest point in the nucleus was offset from the center of the
bulge by $\sim 0\farcs 4$.  The HST Wide-field and Planetary Camera
(WFPC1) later resolved the nucleus into two brightness peaks (Lauer et al. 
1993, hereafter L93), as did more recent HST images taken with WFPC2
(Lauer et al.  1998, hereafter L98).  The optically brighter peak, P1, is
offset $0\farcs 49$ from the bulge photometric center, which coincides
with the fainter peak, P2.  P1 and P2 have central $V$ band surface
brightnesses of $13.4 \magarc$ and $13.7 \magarc$, respectively, when
averaged over a $0\farcs22$ wide slit (L93).  P1 is compact, with a
major-axis core radius of $\sim 0\farcs4$; P2 has a weak stellar cusp,
unlike P1 (L93). 

Near-IR (Mould et al. 1989, Davidge et al. 1997, Corbin et al. 2001),
optical (L93, L98), and far-UV (King et al. 1995, hereafter K95) images
all show that the asymmetric or double-peaked structure of the nucleus is
not caused by dust absorption.  Along with absorption-line strengths from
long-slit spectra (Kormendy \& Bender 1999, hereafter KB99), they also
demonstrate that P1 has a similar stellar content as the rest of the
nucleus, which is unlike any globular cluster or dwarf elliptical.  Thus,
P1 is an intrinsic part of the nucleus, and not an interloping star
cluster (K95).  The $V-I$ color of the nucleus is not the same as that of
the bulge, implying a difference in stellar populations (L98, Bacon et al.
2001, hereafter B01); line strengths in KB99 also show the same.  The
color difference is not agreed upon, however; L98 find that the nucleus is
redder than the bulge, whereas B01 find the opposite.  Sil'chenko et al.
(1998) argue that the nucleus is more metal rich than the bulge, and,
using $H_\beta$ lines to disentangle metallicity and age, find that the
nucleus is a factor of three younger than the bulge. 

\begin{figure}
\plotone{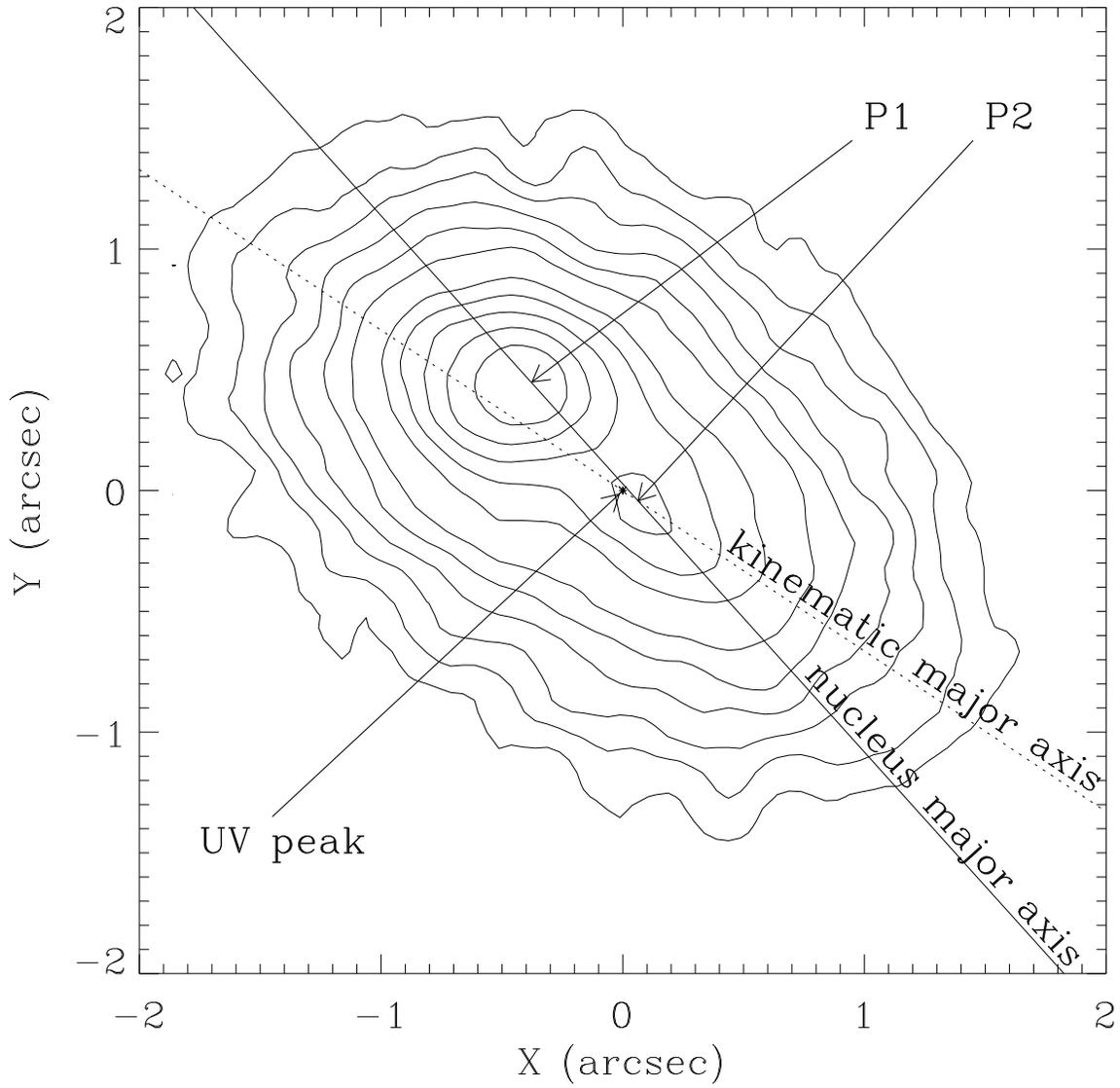}
\caption{\footnotesize
WFPC2/HST photometry in the $I$ (F814W) band from L98.  The image has been 
boxcar
smoothed for clarity.  North points toward the top of the page, while East
points to the left.  Arrows show the locations of P1, P2, and the UV peak; 
the center of the UV peak is denoted by an asterisk at the origin.  The
solid line shows the P1-P2 line ({\it PA}$_d=42\arcdeg$), or the major axis
of the nucleus.  The dotted line shows the kinematic major axis, which is
the line joining the velocity extrema in the two-dimensional kinematic map
({\it PA}$_K=56.4\arcdeg$). 
}
\label{f.nucleus}
\end{figure}

P2 is brighter than P1 in the UV, as a result of an embedded UV-bright
source (hereafter the UV peak; K95, Brown et al. 1998, L98) whose center
is located $0\farcs 076$ toward P1 from the center of P2 in the $I$-band
(B01).  The UV peak is resolved, with a half-power radius of $\sim
0\farcs2$ (Brown et al.  1998, L98, B01).  Brown et al. (1998) show that
the UV peak is consistent with being comprised of extreme horizontal
branch stars with masses between $0.47$ and $0.53$ \msun, but not
consistent with a majority contribution from main sequence stars, blue
stragglers, or post-asymptotic giant branch stars more massive than $0.56
\msun$.  Unpublished spectra from STIS also suggest that the UV peak is
dominated by starlight (E. Emsellem, private communication), rather than a
low-level AGN (K95).  The UV peak is thought to be the location of the
photometric center of the bulge, and the supermassive BH (K95, KB99, B01;
Peng 2002, hereafter P02).  Hereafter in this paper, ``the nucleus''
refers to P1 and P2 together, but does not include the UV peak, which is a
separate nuclear star cluster.  The locations of P1, P2, and the UV peak
with respect to the nucleus as a whole are shown in Figure
\ref{f.nucleus}. 

Groundbased observations at $\sim 1\arcsec$ (FWHM) resolution by Dressler
(1984), Kormendy (1988), Dressler \& Richstone (1988), and van der Marel
et al. (1994) were the first to show that the stellar component in M31's
nucleus rotates rapidly, and that there is a significant velocity
dispersion peak (hereafter the dispersion spike) in the central few
parsecs, both possibly indicating the presence of a central BH of mass
$\sim 10^7 \msun$.  The data show the dispersion spike to be centered
$\sim 0\farcs 6$ away from the peak in brightness (P1) in the Stratoscope
II photometry, and the nucleus to be colder than the bulge on both sides
of the dynamical center.  Two-dimensional kinematic maps obtained by Bacon
et al. (1994) at similar resolution ($0\farcs 87$ FWHM), using the TIGER
integral field spectrograph on the Canada-France-Hawaii Telescope (CFHT),
are consistent with most of the earlier observations;  however, the
dispersion spike in that data set is located $\sim 0\farcs 7$ from P2 on
the anti-P1 side of the nucleus, which places it $\sim 0\farcs 6$ farther
away from P1 than found previously.  The deconvolved TIGER rotation curve
is asymmetric about the rotation center, which is near P2; the maximum
amplitude on the anti-P1 side is roughly $60 \kms$ greater than that on
the P1 side. 

Observations at better spatial resolution ($0\farcs 64$ FWHM) and higher
signal-to-noise (S/N), taken with the Subarcsecond Imaging Spectrograph
(SIS) on CFHT (KB99), show the dispersion spike offset from the UV peak by
$\sim 0\farcs 2$, roughly $0\farcs 4$ less than the Bacon et al. (1994)
offset;  the spike's amplitude is $248 \pm 5 \kms$ before bulge
subtraction and $287 \pm 9 \kms$ after.  The nucleus is cold on both sides
of the UV peak, as in Kormendy (1988);  for example, the dispersion at
$r=0\farcs92$ from the UV peak on the P1 side is $123 \pm 2 \kms$ with the
bulge, and $\sim 100 \kms$ without.  KB99 find that the bulge-subtracted
maximum rotation velocity is $-236 \pm 4 \kms$ on the anti-P1 side, but
only $179 \pm 2 \kms$ on the P1 side, confirming the $\sim 60 \kms$
rotation amplitude asymmetry of Bacon et al. (1994).  When the bulge is
added, the asymmetry is only $\sim 7 \kms$, with a maximum velocity on the
P1 side of $152 \pm 3 \kms$.  The zero velocity crossing is displaced from
the UV peak toward P1 by $0\farcs 051 \pm 0\farcs 014$.  Slit-averaged
velocity profiles from the OASIS integral field spectrograph on CFHT
(B01), which has about twice the spatial resolution ($\sim 0\farcs 4$ -
$0\farcs 5$ FWHM) as TIGER, are consistent with the SIS observations.  B01
measure the kinematic major axis, or the line joining the velocity extrema
in the two-dimensional map, to be at position angle {\it PA}$_K =
56.4\arcdeg \pm 0.2\arcdeg$, which is not on the P1-P2 line ({\it PA}$_d =
42\arcdeg$; see Figure \ref{f.nucleus}). 

The kinematic observations with the best resolution to date ($\sim 0\farcs
1$ FWHM) come from the f/48 long-slit spectrograph of the HST Faint Object
Camera (FOC; Statler et al. 1999).  The rotation curve is resolved through
the rotation center with a projected velocity gradient of $\sim 300
\kmspc$;  the zero velocity crossing is offset by $0\farcs 16 \pm 0\farcs
05$ from P2 towards P1, or $\sim 0\farcs 135$ from the UV peak (using the
spatial registration suggested by B01).\footnote{B01 determined that
spatial shifts must be applied to the FOC and SIS data to register them to
the center of the UV peak in the F300W band, which is the reference center
for the OASIS and STIS data.  The origin defined in Statler et al. (1999)
must be shifted $0\farcs 25$ toward P1, whereas the origin defined in KB99
must be shifted $0\farcs 031$ away from P1.  They also found that a
positive shift of $30 \kms$ must be applied to the FOC velocity profile
for consistency with the STIS profile;  this amounts to adding $30 \kms$
to the systemic velocity.} The rotation curve is asymmetric, as in the SIS
data, with an amplitude asymmetry of at least $60 \kms$ and a P1-side
maximum of $\sim 240 \kms$.  The dispersion spike has amplitude $440 \pm
70 \kms$, but is only offset from P2 by $0\farcs 06$, in contrast to the
$\sim 0\farcs 2$ offset found with SIS and OASIS.  High-resolution
kinematic data from STIS/HST (B01) also show similar asymmetries.  The
rotation curve has an amplitude asymmetry possibly as high as $\sim 90
\kms$, with a maximum rotation amplitude on the P1-side of $201 \pm 5
\kms$.  The velocity gradient is $\sim 220 \kmspc$ through the zero
velocity crossing, which occurs $0\farcs 09$ from the UV peak;  both of
these values are lower than for FOC.  The dispersion spike has amplitude
$321 \pm 33 \kms$, and is located $0\farcs 235$ from the UV peak on the
anti-P1 side.  B01's STIS dispersion spike is substantially more offset
than that in the FOC data.  We refer the reader forward to figures in 
Section \ref{s.bestfits} to see FOC, STIS, SIS and OASIS kinematic profiles. 

Two hypotheses have been explored to account for the photometric and
kinematic asymmetries observed in M31's nucleus:  first, that P1
represents a captured star cluster orbiting around a stellar disk and
central BH (Emsellem \& Combes 1997); second, that P1 is produced by orbit
crowding at apoapsis in an eccentric disk of stars on apse-aligned Kepler
orbits about a BH at P2 (Tremaine 1995, hereafter T95).  Of the two
hypotheses, the evidence strongly points toward the second as
correct.  The eccentric disk picture naturally explains the nearly uniform
colors of the nucleus, since P1 and P2 are the same stellar population. 
It is difficult to explain the colors with the orbiting cluster picture,
since the colors of P1 are unlike any globular cluster or dwarf
elliptical.  A further strike against the cluster picture is demonstrated
in self-consistent N-body simulations by Emsellem \& Combes (1997);  they
find that the timescale for disruption is only $\sim 10^5$ years, so it is
unlikely that such a configuration would be observed.

T95's original model for an eccentric nuclear disk in M31 consists of
three nested and aligned Keplerian ringlets with outwardly decreasing
eccentricities.  Random velocities are roughly accounted for by
convolution with a Gaussian point spread function in the plane of the sky. 
The model fits the photometry of L93 and is broadly consistent with the
ground-based kinematics of Kormendy (1988) and Bacon et al. (1994). 
Though simple, the model predicts many of the asymmetries seen in the more
recent kinematic profiles from SIS, FOC, STIS, and OASIS, including the
displaced rotation center, the asymmetric rotation amplitudes, the low
velocity dispersion at $r \sim 1\arcsec$ on the P1 side of the nucleus,
and the presence of a dispersion spike near P2. 

In its original form, the T95 model is too limited to be used to
constrain the mass of the central BH in M31.  The model ignores
self-gravity, which is necessary to maintain apse-alignment in the disk
against differential precession (T95; Statler 1999, hereafter S99).  Also,
the model does not include a realistic treatment of velocity disperison,
which is needed for an accurate prediction of the dispersion profile. 
Both of these ingredients need to be included self-consistently. 

Hints at how such a model can be constructed were first given by Sridhar
\& Touma (1999).  They compute orbits in nearly-Keplerian potentials with
lopsided perturbations and find a family of periodic loop orbits elongated
in the same sense as the perturbation.  They suggest that the nearly
elliptical periodic parents of such orbits can be used as the backbone
around which an eccentric disk with self-gravity and finite dispersion can
be built.  S99 computes periodic loop orbits for a continuous, uniformly
precessing T95-like disk model, and shows that the requirement of uniform
precession has important consequences for the disk structure.  He finds
that the periodic orbits follow a non-monotonic radial eccentricity
distribution, in which a steep negative eccentricity gradient though the
densest part of the disk is followed by a reversal of the arrangement of
pericenter and apocenter with respect to the BH.  S99 suggests that
approximate self-consistent equilibria can be constructed around such a
sequence of numerically integrated closed periodic orbits, by
approximating a distribution of quasi-periodic orbits about a given
periodic parent with a distribution of Kepler orbits dispersed in
eccentricity and orientation.  Salow \& Statler (2001) use this
approximation to construct radially truncated models that reproduce many
of the features seen in FOC kinematics and one-dimensional HST photometry
within $0\farcs 5$ of the UV peak; the models are built by iteration with
a phase space distribution function (DF) written in terms of the integrals
of motion in the Kepler problem (S99).  They find that the backbone orbits
follow an eccentricity distribution similar to that in S99, which gives
rise to distinctive multi-peaked LOSVDs near the UV peak. 

Several authors have constructed self-consistent eccentric disk models by
other methods.  Jalali \& Rafiee (2001) construct integrable models whose
potentials are of the St\"{a}ckel form in elliptic coordinates.  They show
that models with double nuclei are sustained by four general types of
regular orbits (butterflies, nucleophilic bananas, horseshoes, and aligned
loops).  Their models, however, require that both P1 and P2 have density
cusps, which is not seen in the data.  B01 and Jacobs \& Sellwood (2001)
perform N-body simulations of lopsided ($m=1$) modes in a cold disk
orbiting a central BH, and are able to find models that reproduce some of
the observed features of the nucleus.  More importantly, they demonstrate
that lopsided stellar disks can be long-lived, giving further support to
the eccentric disk picture.  Sambhus \& Sridhar (2002, hereafter SS02)
construct models using a Schwarzschild-type method (Schwarzschild 1979)
with an orbit library composed of both prograde and retrograde orbits. 
They find that the latter are needed to better fit the kinematics and
photometry near P2;  Touma (2002) argues that a small percentage of
retrograde orbits is all that is needed for a Keplerian disk to grow an
unstable lopsided mode.  Both SS02 and B01 find an eccentricity
distribution different than that found by S99 and Salow \& Statler (2001). 
Orbits follow a steep negative eccentricity gradient through the dense
part of the disk, but do not switch their apoapses to the anti-P1 side of
the disk afterward. 

Peiris \& Tremaine (2003, hereafter PT03) have recently shown how a
T95-like model can be extended to three-dimensions.  They construct models
comprised of non-interacting Kepler orbits in the gravitational field of
the BH.  They draw orbital elements from a Monte-Carlo scheme, and
populate the disk with a parametric DF;  orbits are dispersed in
eccentricity, orientation, and inclination, rather than just the first
two, as in Salow \& Statler (2001).  Their models are able to reproduce
most of the important features in HST photometry and SIS and unpublished
STIS (Bender et al.  2003) kinematics within $\sim 1\arcsec$ of the UV
peak.  However, their models are missing self-gravity and gravity-induced
precession in the disk. 

In this paper we extend the self-gravitating, finite dispersion models of
Salow \& Statler (2001) to include a greater radial extent, in order to
rigorously model the double nucleus of M31.  Along with an optimization
routine, these models are used to fit FOC, STIS, and SIS one-dimensional
kinematics, OASIS two-dimensional kinematics, and one and two-dimensional
WFPC2/HST photometry.  Best-fit disk parameters and BH masses are found for a
grid of 24 models by minimizing a chi-square merit function which assesses
agreement between model and data.  The primary result of this paper is an
accurate mass for the BH in M31.  Secondarily, we present the properties 
of the disk that best fits the nucleus.

The plan of this paper is as follows.  In Section \ref{sec:CHAP2}, we give
the details of model construction.  We then provide a description of the
necessary assumptions and instrument specifications needed to find models
that best fit data from the nucleus of M31 in Section \ref{sec:CHAP3}.  In
Section \ref{sec:CHAP4} we present results from a grid of 24 best-fit
models for M31's nucleus, including the BH mass in M31 and disk parameters
and properties.  Section \ref{sec:CHAP5} discusses the connection with
other work.  Finally, Section \ref{sec:CHAP6} presents some brief
concluding remarks. 


\section{Eccentric Disk Models}
\label{sec:CHAP2}

\subsection{Theoretical Basis}
\label{s.theory}

Following Sridhar \& Touma (1999) and S99, we construct
realistic models from sets of quasi-periodic orbits whose parents are
closed periodic loops elongated in the same sense as the lopsided
perturbation.  The parent loops form the ``backbone'' of the disk, around
which quasi-periodic orbits will be populated.  These backbone orbits will
precess and deform under the influence of the disk's self-gravity. 
However, if the mass of the disk is small enough the backbone orbits will
be nearly Kepler ellipses in the rotating frame.  This fact, together with
results from simple orbit integrations in lopsided potentials, is
suggestive of a way to approximate distributions of quasi-periodic orbits
about a given backbone orbit.  Explorations of the orbital structure in a
nearly Keplerian potential perturbed by a slowly precessing eccentric disk
show that quasi-periodic orbits fill bands surrounding the backbone orbits
(Statler \& Salow 2000);  alternatively, they can be thought of as librating 
about the backbone orbits in eccentricity and orientation.  A natural
approximation is then to describe a distribution of quasi-periodic orbits
about a given backbone orbit by a distribution of Kepler orbits dispersed
in eccentricity and orientation.  We follow this approximation, taking it
as a postulate. 

To represent a distribution of Kepler orbits about a given backbone
orbit we use a phase space distribution function (DF) written in terms of
integrals of motion in the unperturbed Kepler potential, $f(a,e,\omega)$,
where $a$ is the semimajor axis, $e$ is the eccentricity, and $\omega$ is
the argument of pericenter of a Kepler orbit;  i.e., $\omega$ is
the direction of the Runge-Lenz vector.  We have chosen a simple DF which
is separable in all three variables;  that is, $f(a,e,\omega) = F(a) F(e)
F(\omega)$.  The details of the DF are given in Section \ref{s.df}. 

Models include a two-dimensional eccentric stellar disk surrounding a
BH of mass $M_{BH}$.  The density distribution of the
disk is fixed in a frame rotating at constant angular speed $\Omega$ about
the center of mass of the system, and is normalized to a total mass
$m=\epsilon M_{BH}$.  The black hole is located at the origin of a
Cartesian coordinate system, and the disk is oriented such that its
major-axis lies along the $x$ coordinate line.  Models are computed on a
$200 \times 200$ grid with spacing $l=0.25$ in dimensionless units where
$G=M_{BH}=1$.  The potential of a spheroidal bulge component is not
included, since its effect on the precession frequencies of Kepler orbits
in the absence of the disk potential is small.\footnote{Bulge-induced
precession frequencies are less than 10\% of $\Omega$ for a
class of spherical, nonrotating $\eta$-models to be discussed in Section 
\ref{s.getobservables}.}

\subsection{The Distribution Function}
\label{s.df}

$F(e)$ and $F(\omega)$ together provide our prescription for
the way dispersed Kepler ellipses are distributed about the sequence of
backbone orbits.  We have considered two versions of $F(e)$.  The
first is a Gaussian distribution of eccentricities given by

\begin{equation}
F(e) = \exp \left[ - {[e-e_0(a)]^2 \over 2 {\sigma_e}^2} \right],
\label{e.fofeg}
\end{equation}

\noindent
where $e_0(a)$ describes the sequence of backbone orbits, and the constant
$\sigma_e$ determines the spread in eccentricity about a given backbone
orbit.  The second 
version of $F(e)$ is a Rayleigh distribution of 
eccentricities, and is given by

\begin{equation}
F(e) = |e| \exp \left[ - {[e-e_0(a)]^2 \over 2 {\sigma_e}^2} \right],
\label{e.fofer}
\end{equation}

\noindent
where $e_0(a)$ and $\sigma_e$ have the same meaning as for the Gaussian 
distribution.  The velocity distribution for the Gaussian form of $F(e)$ is 
singular at 
$e=0$, and thus somewhat unphysical.  As a result, an extra population of 
circular orbits will be populated, in addition to the normal eccentric orbit 
population around $e_0(a)$.  The Rayleigh form adds an extra factor of 
$e$ to 
ensure finiteness at $e=0$.  For both forms of $F(e)$ we use a Gaussian 
distribution of orientations. F($\omega$) is given by

\begin{equation}
F(\omega) = \exp \left[ - {\omega^2 \over 2 {\sigma_\omega}^2} \right],
\label{e.fofg}
\end{equation}

\noindent
where the constant $\sigma_\omega$ is the dispersion in $\omega$.

The function $F(a)$ gives the mass per unit interval of semimajor axis,
and thus controls the radial mass distribution.  We have chosen a form for
$F(a)$ which allows variability in the strength of the central density
minimum, and in the strength and width of the maximum peak in the mass
distribution.  $F(a)$ is described by two functions joined together,
$F_I(a)$ and $F_O(a)$, which represent the inner and outer parts of the 
disk, respectively.  We use

\begin{equation}
F(a) = \left\{ \begin{array}
	{l@{\quad:\quad}r}
	F_I(a) & a \leq a_{max}\\
	F_O(a) & a > a_{max},
	\end{array} \right.
\label{e.fofa}
\end{equation}

\noindent
where $a_{max}$ is the value of $a$ at which $F_I(a)$ is maximum, 
$F_{max}$.  $F_I(a)$ is given by

\begin{equation}
F_I(a) = \mbox{\rm max}(a-\Delta,0)\exp \left[ - {(a-a_0)^2 \over 2 
{\sigma_a}^2} \right],
\label{e.fofai}
\end{equation}

\noindent
where $\sigma_a$ controls the width of the inner density distribution,
$\Delta$ determines the strength of the central density minimum, and $a_0$
sets the length scale;  we set $a_0=2$.  $F_O(a)$ is given by

\begin{equation}
F_O(a) = C F_I(a) + (1 - C) F_{max}\sech \left[ (a-a_{max}) \over \sigma_r 
\right].
\label{e.fofao}
\end{equation}

\begin{figure}
\plotone{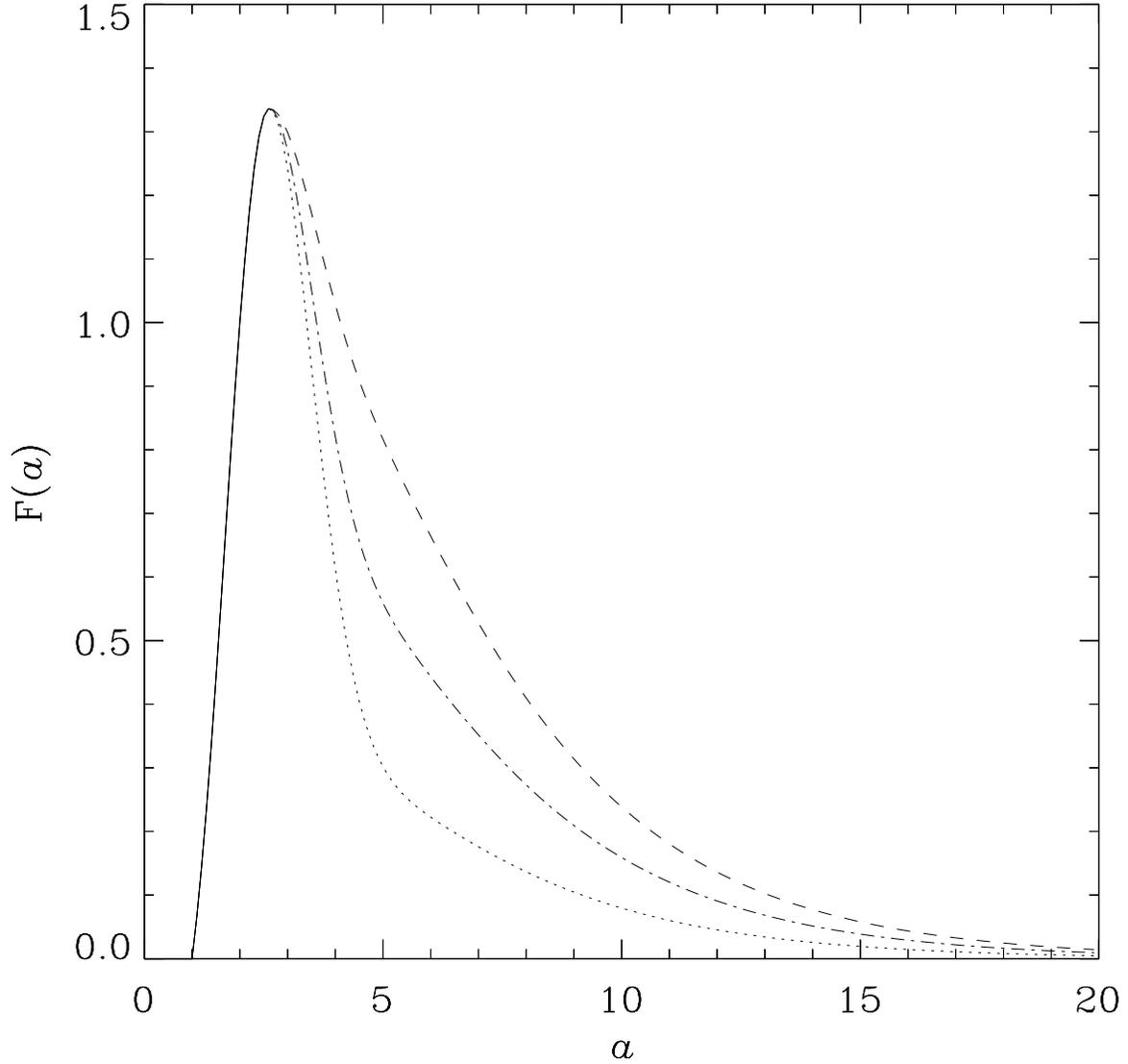}
\caption{\footnotesize
The function $F(a)$, which controls the radial mass distribution. $F(a)$
is shown in model units for three values of the parameter $C$ in Equation
\ref{e.fofao}, which determines how mass is distributed between the inner
and outer parts of the disk.  $C$ also partially determines how fast the
density drops-off away from maxiumum.  The solid line shows $F_I(a)$ for
$\sigma_a=1$, $\Delta=1$, and $a_0=2$.  The dotted, dash-dotted, and
dashed lines show $F_O(a)$ when $C$ has value $0.75$, $0.5$, and $0.25$,
respectively.  A larger value of $C$ results in a weaker outer disk and a
steeper drop-off in density outside of maximum. 
}
\label{f.fofa}
\end{figure}

\noindent
The constant $C$ has two effects on the behavior of the disk:  First, it 
determines how much mass is distributed to the outer part of the disk, 
and second, it partially determines how quickly the density drops-off 
away from maximum for $a \geq a_{max}$.  Larger values of $C$ result in 
weaker outer disks and steeper drop-offs in density outside of maximum.  
Figure \ref{f.fofa} shows the behavior of $F(a)$ for three values of $C$.

The parameter $\sigma_r$ is used to extend the disk to the
desired cutoff radius, $R_d$.  Simple algebra shows that if
$F_O(a)=\alpha$ at $a=R_d$, where $\alpha$ is some small number, then
$\sigma_r$ is given by

\begin{equation}
\sigma_r = {(R_d - a_{max}) \over \ln ({(1 - C) F_{max} \over \alpha} + 
\sqrt{[{(1 - C) F_{max} \over \alpha}]^2 - 1})}.
\label{e.sigmar}
\end{equation}

\noindent
We set $\alpha = {F_{max} / 100}$ to ensure sufficiently small densities 
at $a=R_d$.

\subsection{The Construction Scheme}
\label{s.construction}

A model is specified by the parameters $\epsilon$, $\Omega$, $\sigma_e$,
$\sigma_\omega$, $\sigma_a$, $\Delta$, $C$, $a_0$, and $R_d$.  Once these
are given, an initial guess for $e_0(a)$ must be provided.  We choose
$e_0(a)={1 \over 2}(1 - {a \over R_d})$, which gives an initial density
maximum at apoapsis.  This choice was made because it leads to rapid
convergence, but the results are insensitive to the initial guess. 
Following specification of $e_0(a)$, construction proceeds iteratively 
(see Salow \& Statler 2001). 

Construction begins by expressing the DF in terms of position and 
velocity using the standard Keplerian relations:

\begin{equation}
a=-{1 \over {2 E}}
\label{e.kra}
\end{equation}
\begin{equation}
e=\sqrt{1 + 2 E h^2}
\label{e.kre}
\end{equation}
\begin{equation}
\omega=\arctan{{A_y} \over {A_x}}
\label{e.krg}
\end{equation}

\noindent
where $E={1 \over 2}({v_x}^2 + {v_y}^2) + \Phi$ is the energy per unit 
mass, $\Phi= - (\sqrt{x^2 + y^2})^{-1}$ is the unperturbed potential, and 
$h= x v_y - y v_x$ is the angular momentum per unit mass (Murray \& 
Dermott 1999).  The quantities
$A_x = v_y h + x \Phi$ and $A_y = - (v_x h) + y \Phi$ are the $x$ and $y$
components of the Runge-Lenz vector, respectively.  To avoid
discontinuities in the DF, we allow $e$ to be negative and define $\omega$
to lie between $\pm {\pi/2}$. 

The disk density $\rho (x,y)$ is found by integrating the DF over velocity
at each grid point, and then normalizing the grid to total disk mass $m$. 
The potential of the disk is computed using Fast Fourier Transforms (FFTs)
and the discrete fourier convolution theorem (see Section 2.8 of Binney \&
Tremaine 1987).  Zero padding is used to suppress Fourier images.  We use
a softened point-mass kernel of one grid spacing for the Green function. 
The disk potential is added to the potential of the black hole to form the
total potential.  The total potential is rotated at frequency $\Omega$
about the center of mass to include inertial effects from the rotating
frame;  only Coriolis forces are included, since centrifugal terms are of
order $\Omega^2$. 

Numerical integration of the equations of motion in the rotating frame is
performed to find the set of closed periodic orbits that circulate in the
prograde direction and precess uniformly in the total potential.  Orbits
are initially launched perpendicularly from the $x$-axis, and the
velocities are varied until the next $x$-axis crossing occurs with $v_x=0$
(S99).  These will be the backbone orbits for the next iteration. 
To ensure that only nearly-Keplerian orbits are found, the total period
and computed semiminor axis have to be within $50\%$ and $20\%$,
respectively, of those values for a Kepler orbit with the same semimajor
axis and eccentricity.  These two conditions enable separation of
higher-order resonant orbits from nearly-Keplerian orbits for a wide 
range of tested parameter values. 

The backbone orbits are expressed as a new function $e_0(a)$.  This is
done by noting the positions where the orbit crosses the $x$-axis at
positive ($x_+$) and negative ($x_-$) values of $x$.  Following S99, $e$
and $a$ are determined using $e \equiv (x_- + x_+)/(x_- - x_+)$ and $a
\equiv (x_+ - x_-)/2$.  In some cases this is all that must be done; 
however, for large values of $\Omega$ the sequence of periodic orbits
truncates inside $R_d$.  When this occurs, $e_0(a)$ is extended out to
$a=R_d$ using a function chosen to mimic the behavior of $e_0(a)$ for
models with no truncation.  Details of this are given in Appendix
\ref{App_2}.  After extending $e_0(a)$, the quantity $\sigma_r$ in
Equation \ref{e.fofao} is updated to ensure that the disk extends out to
$R_d$, since the physical length scale can change at each iteration (see
Section \ref{s.scalezero}).  A new density distribution is then found and
the aforementioned sequence continues.  Iterations continue until the
fractional change in the density per iteration is less than 5\% everywhere
and less than 1\% on average. 

\subsection{Projecting the Model}
\label{s.getobservables}

\begin{figure}
\plotone{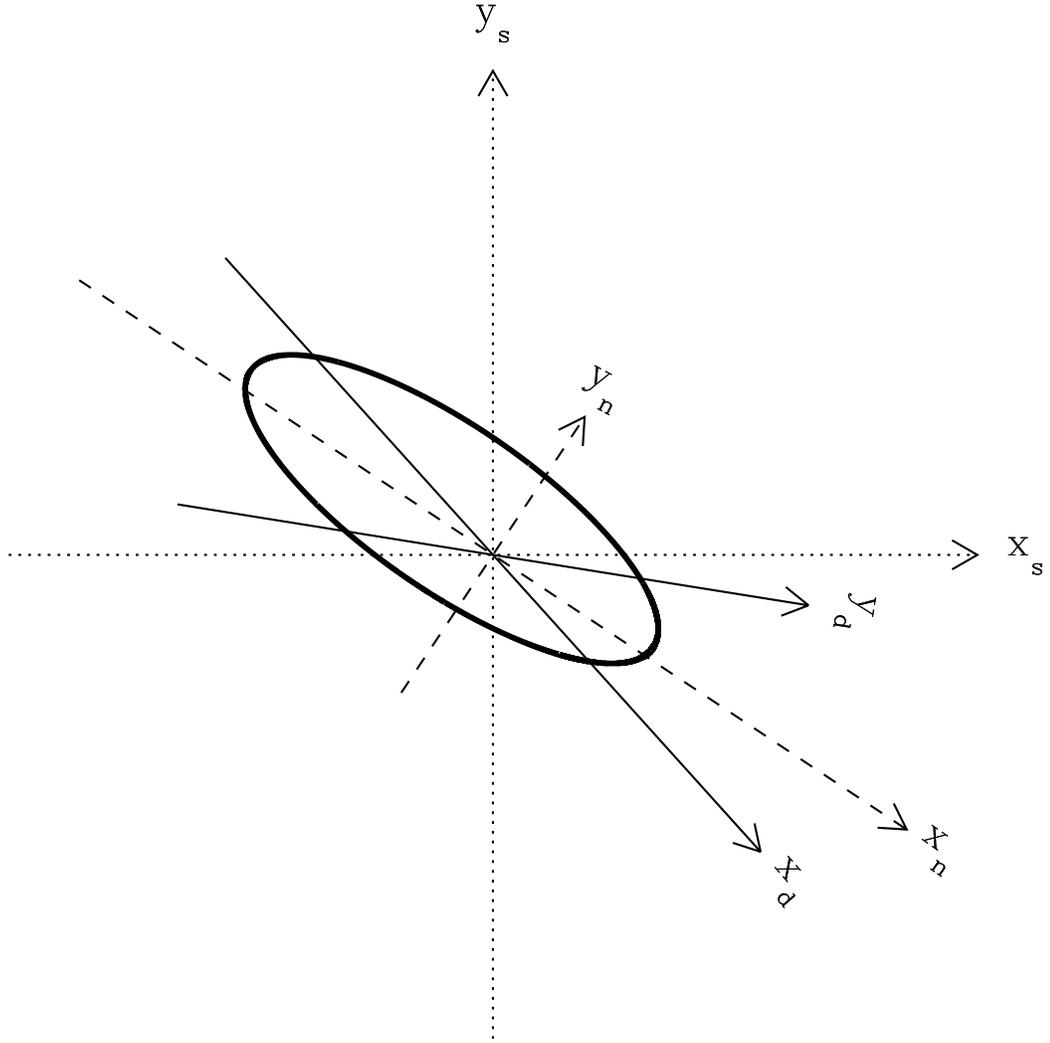}
\caption{\footnotesize
The three coordinate systems used to construct and project a model, as
seen on the plane of the sky for a disk (shown as an ellipse with $e=0.3$)
inclined at $i=70 \arcdeg$ with the line of nodes at {\it PA}$_n=56.4
\arcdeg$ and the disk major axis at {\it PA}$_d=42\arcdeg$.  All 
coordinate axes have the same unprojected length.  Velocity
moments are projected onto $(x_s,y_s)$ in the sky plane, with $y_s$
pointing North ({\it PA}=0\arcdeg) and $x_s$ pointing West.  $(x_d,y_d)$
is the system in which the disk is constructed, with its major axis along
$x_d$.  $(x_n,y_n)$ is oriented such that $x_n$ lies along the line of
nodes and $y_n$ is in the plane of the disk.
}
\label{f.coords}
\end{figure}

We project two-dimensional models onto the plane of the sky.  Along with
inclination ($i$), two position angles in the plane of the sky must be
specified to fully determine a disk's orientation:  the position angle of
the major axis of the disk ({\it PA}$_d$) and the position angle of the
line of nodes ({\it PA}$_n$).\footnote{Positive position angles are
measured eastward from North on the plane of the sky.} We refer to three
coordinate systems to describe how models are projected onto the sky.  The
first, $(x_s,y_s)$, is a system on the plane of the sky for which $y_s$
points along position angle {\it PA}$=0 \arcdeg$.  The second,
$(x_d,y_d)$, is the system in which the disk lies with its major axis
along $x_d$.  The third system, $(x_n,y_n)$, is oriented such that $x_n$
lies along the line of nodes and $y_n$ is in the plane of the disk. 
Figure \ref{f.coords} shows the relationship between these three
coordinate systems as seen on the plane of the sky. 

For computational efficiency we perform kinematical modeling using
velocity moments, rather than the full LOSVD.  We do, however, find LOSVDs
for minimized models at specific locations, as described below in Section
\ref{s.losvds}. 

Moments of the LOSVD are found on a $200 \times 200$ grid with spacing $0
\farcs 02$ in the $(x_s,y_s)$ system.  The zeroth moment, $\rho$, is found
by projecting the density distribution onto the sky directly.  The first
and second moments, $\rho v$ and $\rho v^2$, are found using the DF for
the converged model.  Each point in the grid is transformed to a point in
the disk plane, $(x_d,y_d)$.  At $(x_d,y_d)$ the DF and the Kepler
relations (Equations \ref{e.kra}, \ref{e.kre}, and \ref{e.krg}) are used
to get the distribution of velocities $f({v_x}_d,{v_y}_d)$ on a $200
\times 200$ velocity grid with spacing $0.075$ in units where $G=M_{BH}=1$
and $a_0=2$.  The distribution $f({v_x}_d,{v_y}_d)$ is then transformed to
the $(x_n,y_n)$ coordinate system to obtain the distribution
$f({v_x}_n,{v_y}_n)$ on a similar velocity grid.  This distribution is
transformed to an inertial frame to include the $\vec{\Omega} \times
\vec{r}$ contribution to the velocity, and integrated over ${v_x}_n$ to
give $f({v_y}_n)$, the unprojected disk-plane LOSVD.  Multiplying
$f({v_y}_n)$ by the projection factor $\sin i$ and scaling to physical
velocities gives the LOSVD on the sky at point $(x_s,y_s)$.  The moments
$\rho v$ and $\rho v^2$ are obtained from this LOSVD by one-dimensional
numerical integration.  Moments from the bulge (see below) are then added
to those of the disk, giving the three projected moment distributions
$\rho$, $\rho v$, and $\rho v^2$ on the sky grid. 

Moment distributions are convolved with appropriate spatial point-spread
functions (PSFs) for the observing instruments.  The convolved grids are
then observed over a slit to obtain one-dimensional kinematics or
photometry, or binned for two-dimensional observations.  One-dimensional
observations are made by averaging over a slit of width $w$ and pixel
scale $l$ at a given position angle {\it PA}.  Two-dimensional
observations are obtained by averaging over a square pixel of scale $l$. 
Averaged moments yield line-of-sight rotation ($v$), velocity dispersion
($\sigma$), and surface brightness ($\mu$) profiles.  We follow the usual
convention that objects moving away from the observer have positive
velocities. 

To ensure proper functioning of our code, we generated kinematic and
photometric profiles using the distribution function for a Keplerian disk
with constant surface density and a Rayleigh distribution of
eccentricities;  a Rayleigh distribution is equivalent to a Schwarzschild
distribution in velocity (Dones \& Tremaine 1993).  These profiles were
compared with similar profiles generated from analytically determined
velocity moments of the distribution.  Moments were projected onto the
sky, convolved by numerical integration with a PSF, and then observed over
a slit for the comparison.  Close agreement was found for the FOC, STIS, 
and SIS slits at numerous position angles (see Section \ref{sec:CHAP3} 
for instrument specifications). 

The bulge and central cusp are approximated by a spherical, non-rotating
$\eta$-model that dynamically includes the influence of the BH (Tremaine
et al. 1994).  An $\eta$ model is specified by parameters $M_{BH}$ and
$\eta$, where $\eta$ determines the central cusp strength;  the models
have outer density profiles with $\rho \propto r^{-4}$ and central
power-law density cusps with $\rho \propto r^{3-\eta}$ for $0 < \eta \leq
3$.  The  bulge model is expressed in physical units by two additional 
parameters, $M_b$ and $r_0$, which represent the total bulge mass and 
scale length, respectively.  The bulge is always centered on the BH.

Scaling to physical units requires specification of the mass-to-light
ratio ($M/L = \Upsilon$) and the distance ($D$) to the nuclear disk's host
galaxy. 

\subsection{LOSVDs}
\label{s.losvds}

LOSVDs can change significantly over small spatial scales, so it is
necessary to use a finer grid for their computation and convolution.  We
find the full LOSVD at any given point on the sky by constructing a $200
\times 200 \times 200$ data cube centered on that point.  The first two
dimensions represent the spatial coordinates, while the third represents
the LOSVD at that point.  We set the spatial length per pixel to $0 \farcs
002$ and bin velocities to an instrument-specific resolution.  The spatial
scale extends beyond $w/2+4\sigma_{I}$, where $\sigma_{I}$ is the width of
the observing instrument's PSF and $w$ is the width of the observing slit. 

The data cube is built following a procedure similar to that described
previously for the moment distributions, with the exception that now the
LOSVD is recorded into the data cube instead of having the velocity
moments calculated.  The contribution of the bulge to the total LOSVD is
described by a Gaussian with the projected dispersion of the appropriate
$\eta$-model.  Convolution is performed by marching through the cube in
velocity and convolving each two-dimensional cut with the PSF.  The
convolved data cube is then averaged over the slit width $w$ and pixel
size $l$. 


\section{Modeling Specifics for M31}
\label{sec:CHAP3}

In this Section we provide details necessary to describe how the
construction technique given in Section \ref{sec:CHAP2} is used to find
models that best fit kinematic and photometric data from M31's
nucleus.  Best-fit parameters are found for a grid of models by minimizing
a chi-square merit function. 

\subsection{Assumptions}
\label{s.assume}

We take {\it PA}$_d = 42 \arcdeg$, the P1-P2 axis measured from WFPC2
photometry (B01).  B01 find that the major axis of the nucleus is close to
the OASIS-measured kinematic axis ({\it PA}$_K = 56.4 \arcdeg$), so we
assume that {\it PA}$_n = 56.4 \arcdeg$.  Disk inclination is either fixed
to $i = 52.5 \arcdeg$ or left as a free parameter (see Section
\ref{s.grid});  the fixed value of $i$ is representative of the
inclination found by deprojecting the nucleus, assuming the disk is cold
and thin with nearly-circular outer isophotes (B01, SS02, P02). 

The disk and bulge are assigned $V$ band mass-to-light ratios $\Upsilon_V
= 5.7$, as found from dynamical modeling (Kormendy 1988, Dressler \&
Richstone 1988) and corroborated by a center-of-mass analysis (KB99). 
This value for $\Upsilon_V$ is identical to that used in other
recent investigations of the nucleus of M31 (e.g., SS02, P02, PT03); like 
the other authors, we take it as a fixed parameter with no uncertainty.  
The stars are given colors $V-I = 1.348$, the value $4\arcsec$ from P2
(Table 3 of L98).  M31 is assumed to be located at a distance $D = 770
\kpc$, based on Cepheids (Freedman \& Madore 1990, Kennicutt et al. 1998),
red giant branch stars and globular clusters (Holland 1988), and red clump
stars with parallaxes (Stanek \& Garnavich 1998);  at this distance $1
\arcsec = 3.733$ pc. 

\subsection{The Length Scale and Zero Point}
\label{s.scalezero}

A model is mapped onto the data by two free parameters: a linear scale
factor, $D_{P1}$, and a sliding offset along the major axis of the disk
({\it PA}$_d=42 \arcdeg$), $D_{P2}$.  $D_{P1}$ gives the separation between
the BH and the center of P1 in arcseconds, while $D_{P2}$ specifies the
separation between the BH and the data origin.  In other words, the BH is
assumed to lie somewhere along the major axis, and its exact location is
determined by the data. 

\subsection{The Data Sample}
\label{s.datasample}

The kinematic data include one-dimensional stellar kinematics from FOC,
STIS, and SIS, and two-dimensional stellar kinematics from OASIS.  We
consider only $v$ and $\sigma$ data falling within $2\arcsec$ of the BH
when fitting.  Within this range are $46$, $58$, and $32$ $v$ and $\sigma$
values for FOC (from Table 1 of Statler et al. 1999), STIS (from the website
address in B01), and SIS (from Table 2 of KB99), respectively.  OASIS data
consists of $1319$ $v$ and $\sigma$ measurements from the high-resolution
``M2'' data set (from the website address in B01). 

The photometric data include both one (1WFPC2) and two-dimensional
(2WFPC2) surface brightness profiles taken from the deconvolved $I$ band
WFPC2/HST image of M31 (L98).  The one-dimensional profile is obtained by
averaging the image over a slit of width $w=0 \farcs 353$ and pixel scale
$l=0 \farcs 0456$ at position angle {\it PA}$=52.5 \arcdeg$, as in KB99. 
At this pixel scale, $88$ data points fall within $2\arcsec$ of the BH. 
The zero point is found by comparing this profile with Figure 8 of KB99,
with a shift of $13.9 \magarc$ applied to Figure 8 to express it in
physical units (J. Kormendy 2001, private communication).  The brightness
profile is converted to the $V$ band using the assumed $V-I=1.348$.  The
two-dimensional profile consists of the $I$ band image binned on a $80
\times 80$ grid with spacing $0 \farcs 05$.  The zero point for the raw
$I$ band image is found by comparison with Table 3 of L98. 

In M31, surface brightness fluctuations completely dominate the noise
statistics of the WFPC2 image (T. Lauer 2002, private communication).  To
estimate errors for fitting purposes we used the IRAF routines ``ellipse''
and ``bmodel'' to make a smooth image, which was then subtracted from the
WFPC2 image to form an artificial ``sigma'' image.  Error estimates for
the one-dimensional profile were obtained by finding the standard
deviation of fluctuations within the area covered by the slit, at each
position along the slit.  Errors for the two-dimensional profile were
found by finding the standard deviation of fluctuations within each bin.

For computational convenience only, the kinematic and photometric data
were shifted to a spatial zero point at the center of P2.  The data were
first registered to the UV peak (as in B01), and then shifted
by the $0 \farcs 076$ P2-UV peak separation along the kinematic axis. 
However, all of the results in this paper are shown relative to a spatial
origin at the UV peak. 

\subsection{Instrument Specifics}
\label{s.instruments}

FOC observations were made with the f/48 long-slit spectrograph at
position angle {\it PA}$=42 \arcdeg$.  The slit has width $w=0 \farcs 063$
and pixel size $l=0 \farcs 028$.  The PSF was modeled using the software
package Tiny Tim version 6.0 (Krist \& Hook 1999) from STScI.\footnote{see
http://www.stsci.edu/software/tinytim for details} A sum of three Gaussian
functions with identical amplitudes was fit to the azimuthally-averaged
PSF profile as an approximation;  the three Gaussians have dispersions
$\sigma_1 = 0 \farcs 0417$, $\sigma_2 = 0 \farcs 0140$, and $\sigma_3 = 0
\farcs 0090$.  STIS observations were made with the G750M first-order
grating at position angle {\it PA}$=39 \arcdeg$.  The slit has width $w=0
\farcs 1$ and pixel size $l=0 \farcs 05$.  The PSF was modeled as the sum
of two round two-dimensional Gaussians with parameters $\sigma_1 = 0
\farcs 03223$, $\sigma_2 = 0 \farcs 130853$, and amplitude ratio $I_2/I_1
= 0.053784$ (E.  Emsellem 2002, private communication).  The G750M grating
has a velocity resolution ($\sigma_v$) of $\sim 38 \kms$, so LOSVDs are
binned to $40 \kms$ (see Section \ref{s.losvds}).  SIS observations were
taken at position angle {\it PA}$=52.5 \arcdeg$ over a slit of width $w=0
\farcs 353$ and pixel scale $l=0 \farcs 0864$.  The PSF is given in
analytic form in Equation 3 of KB99.  OASIS observations were made on a
two-dimensional spectrograph with square pixels approximately $l=0 \farcs
11$ in size.  The PSF for the M2 dataset is given in Table 3 of B01 as the
sum of three Gaussians, with $\sigma_1 = 0 \farcs 15$, $\sigma_2 = 0
\farcs 29$, $\sigma_3 = 0 \farcs 448$, $I_2/I_1 = 0.98$, and $I_3/I_1 =
0.023$. 


\subsection{The Grid of Models}
\label{s.grid}

To quantify some of the possible systematic effects in modeling M31's
nucleus we compute best-fit models for three sets of kinematic and
photometric data.  Data Set 1 includes FOC, STIS, and 1WFPC2.  Data Set 2
adds SIS and OASIS to Data Set 1.  Data Set 3 is identical to Data Set 2,
except that 1WFPC2 is replaced by 2WFPC2;  each kinematical data point in
set 3 is weighted by a factor of three to make the kinematics and
photometry equivalent in the fitting procedure.

\begin{figure}
\plotone{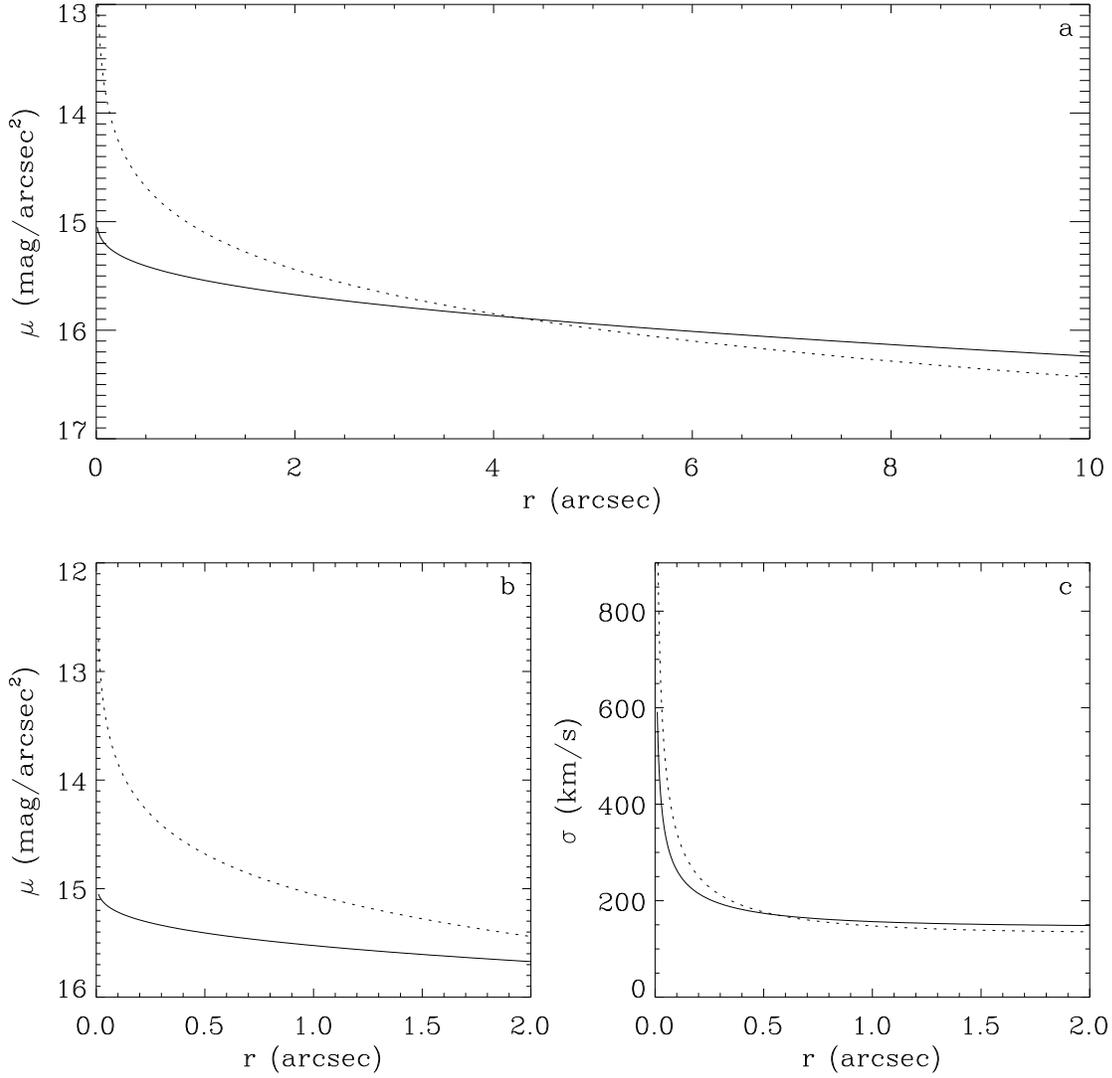}
\caption{\footnotesize
Projected profiles for the two bulge models.  
The bulge is approximated by a spherical, non-rotating $\eta$-model that 
dynamically includes the influence of the BH (Tremaine et al. 1994).
Solid lines show the weak bulge, which resembles B01's multi-Gaussian
expansion model from $4\arcsec$ to $10\arcsec$.  Dotted lines show the
strong bulge, which has a peak projected brightness roughly equivalent to 
that at P2, and the
same brightness as the weak bulge at $r \simeq 4\arcsec$. Panel (a) shows the
surface brightness in the inner $10\arcsec$. 
Panel (b) shows the inner $2\arcsec$ of panel (a).  Panel (c) shows the
projected velocity dispersion. 
}
\label{f.bulges}
\end{figure}

For each Data Set, we compute models for two choices of bulge model,
$F(e)$, and $i$;  thus, there is a group of 8 best-fit models associated
with each Data Set.  The two bulge models are referred to as weak and
strong, and are given by $\eta$-model parameters $\eta = 2.17$, $r_0 =
108.0$, and $M_b = 2.3 \times 10^{10} \msun$ and $\eta = 1.55$, $r_0 =
500.0$, and $M_b = 5.9 \times 10^{10} \msun$, respectively.  The weak
bulge resembles B01's multi-Gaussian expansion model from $4\arcsec$ to
$10\arcsec$.  The strong bulge has a one-dimensional peak projected
surface brightness of $13.65 \magarc$, which is roughly equivalent to that
at P2, and has the same brightness as the weak bulge at $4\arcsec$. 
Figure \ref{f.bulges} shows projected surface brightness and velocity
dispersion profiles for the two bulge models.  The two $F(e)$s are the
Gaussian and Rayleigh distributions given in Equations \ref{e.fofeg} and
\ref{e.fofer}.  The inclination is either set to $i=52.5 \arcdeg$ and
$R_d$ allowed to be a free parameter, or $i$ is free and $R_d$ is fixed
to $R_d=3 \arcsec$. 

\subsection{Chi-square Minimization and Analysis}
\label{s.chimin}

Best-fit models are found using the downhill simplex method (Press et al.
1992) to minimize the reduced chi-square function

\begin{equation}
\chi^2_\nu(\vec{a}) = {1 \over {N-M}} \sum_{i=1}^N \left [{{y_i - y(x_i ; 
\vec{a})} \over \sigma_i} \right ]^2 ,
\label{e.chi2}
\end{equation}

\noindent
where $\vec{a}$ is the set of $M$ fitting parameters, $y_i$ is one of the
$N$ observed data points, $y(x_i ; \vec{a})$ is the modeled data point
corresponding to $y_i$, and $\sigma_i$ is the error estimate associated
with $y_i$.  The minimization is $11$-dimensional, since the parameter set
$\vec{a}$ includes $\epsilon$, $\Omega$, $\sigma_e$, $\sigma_\omega$,
$\sigma_a$, $\Delta$, $C$, $M_{BH}$, $D_{P1}$, $D_{P2}$, and $i$ or $R_d$. 

Formal error estimates in the fitted parameters $\vec{a}$ are obtained by 
first forming the curvature matrix

\begin{equation}
\alpha_{kl} \approx \sum_{i=1}^N {1 \over \sigma_i^2} \left [{{\partial
y(x_i;\vec{a})} \over {\partial a_k}} {{\partial y(x_i;\vec{a})} \over
{\partial a_l}} \right ] ,
\label{e.curvature}
\end{equation}

\noindent
where $a_k$ is the $k^{th}$ parameter.  The partial derivatives are 
made using the central difference formula at the location of the 
minimum $\chi^2$ value.  Second derivative 
terms in Equation \ref{e.curvature} have been ignored, following the
recommendation of Press et al. (1992).  The covariance matrix $[C]$ is
then found by inverting the curvature matrix.  Squared errors in the 
parameters are given by the diagonal elements of $[C]$. 

Initial parameter estimates for the fitting routine were chosen based on
which Data Set the model is fitted against.  For models fitting Data Set
1, the initial parameters were assigned arbitrarily from the ``M31-like''
region of parameter space, as found from trial-and-error searches.  The
results from the sub-grid fitting Data Set 1 were then used as initial
conditions for the corresponding sub-grid fitting Data Set 2.  Similarly,
models from the sub-grid fitting Data Set 2 were used as starting points
for models fitting Data Set 3.


\section{Modeling Results for M31}
\label{sec:CHAP4}

\subsection{Best-Fit Models}
\label{s.bestfits}

Results for the grid of 24 models described in Section \ref{s.grid} are
given in Tables \ref{t.results1}, \ref{t.results2}, and \ref{t.results3}. 
Each table gives fitted parameters expressed in physical units, with
formal errors, for the 8 best-fit models associated with each Data Set. 
The disk mass, $M_d$, is given in place of $\epsilon$, and the peak
eccentricity in $e_0(a)$, $e_{max}$, and reduced chi-square value,
$\chi_\nu^2$, are provided as well.  We now give a brief data-model
comparison for a representative model from each of the three Tables. 

\begin{center}
\begin{deluxetable}{lcccccccc}
\tablecolumns{9}
\tablewidth{0pt}
\rotate
\tabletypesize{\tiny}
\tablecaption{Parameter values for models fitting Data Set 1
\label{t.results1}}
\tablehead{{Model} & \colhead{1} & \colhead{2} & \colhead{3} &
\colhead{4} & \colhead{5} & \colhead{6} & \colhead{7} & \colhead{8} \\[7pt]
{Bulge} & \colhead{Weak} & \colhead{Weak} & \colhead{Strong} &
\colhead{Strong} & \colhead{Weak} & \colhead{Weak} & \colhead{Strong} & 
\colhead{Strong} \\[7pt]
{$F(e)$} & \colhead{Rayleigh} & \colhead{Gauss} & 
\colhead{Rayleigh} &
\colhead{Gauss} & \colhead{Rayleigh} & \colhead{Gauss} & 
\colhead{Rayleigh} & \colhead{Gauss} \\[7pt]
{Inclination} & \colhead{Free} & \colhead{Free} & 
\colhead{Free} &
\colhead{Free} & \colhead{Fixed} & \colhead{Fixed} & 
\colhead{Fixed} & \colhead{Fixed}}
\startdata
$M_{BH}$ ($\times 10^7 \msun$)&6.41 $\pm$ 0.21&5.80 $\pm$ 0.17&6.11 $\pm$ 
0.19&6.09 $\pm$ 0.30&5.51 $\pm$ 0.19&6.18 $\pm$ 0.17&6.15 $\pm$ 0.13&6.05 
$\pm$ 0.14\\[7pt]
$M_d$ ($\times 10^7 \msun$)&1.15 $\pm$ 0.09&1.38 $\pm$ 0.09&1.52 $\pm$ 
0.15&1.77 $\pm$ 0.14&1.62 $\pm$ 0.12&2.01 $\pm$ 0.12&1.61 $\pm$ 0.09&1.70 
$\pm$ 0.10\\[7pt]
$\Omega$ ($\kmspc$)&45.7 $\pm$  8.4&32.5 $\pm$  9.4&41.8 $\pm$ 16.4&43.6 
$\pm$  6.3&44.1 $\pm$ 13.9&31.5 $\pm$ 11.9&43.4 $\pm$ 12.7&36.8 $\pm$ 
16.8\\[7pt]
$\sigma_e$&0.1821 $\pm$ 0.0049&0.2231 $\pm$ 0.0055&0.1840 $\pm$ 
0.0067&0.2403 $\pm$ 0.0078&0.1802 $\pm$ 0.0058&0.2346 $\pm$ 0.0057&0.1794 
$\pm$ 0.0067&0.2410 $\pm$ 0.0080\\[7pt]
$\sigma_{\omega}$ ($rad$)&0.674 $\pm$ 0.065&0.663 $\pm$ 0.037&0.829 $\pm$ 
0.060&0.618 $\pm$ 0.042&0.812 $\pm$ 0.036&0.760 $\pm$ 0.039&0.830 $\pm$ 
0.078&0.600 $\pm$ 0.050\\[7pt]
$\sigma_a$ ($arcsec$)&0.0020 $\pm$ 0.0035&0.0017 $\pm$ 0.0006&0.0017 
$\pm$ 0.0006&0.0001 $\pm$ 0.0020&0.0018 $\pm$ 0.0016&0.0114 $\pm$ 
0.0067&0.0014 $\pm$ 0.0020&0.0074 $\pm$ 0.0020\\[7pt]
$\Delta$ ($arcsec$)&0.14 $\pm$ 0.11&0.23 $\pm$ 0.14&0.15 $\pm$ 0.07&0.05 
$\pm$ 0.23&0.07 $\pm$ 0.12&0.08 $\pm$ 0.04&0.15 $\pm$ 0.04&0.06 $\pm$ 
0.21\\[7pt]
$C$&0.438 $\pm$ 0.097&0.522 $\pm$ 0.067&0.583 $\pm$ 0.084&0.506 $\pm$ 
0.068&0.470 $\pm$ 0.088&0.460 $\pm$ 0.187&0.567 $\pm$ 0.084&0.487 $\pm$ 
0.153\\[7pt]
$R_d$ ($arcsec$)&3.00&3.00&3.00&3.00&3.82 $\pm$ 0.26&3.71 $\pm$ 0.28&3.49 
$\pm$ 0.28&3.32 $\pm$ 0.28\\[7pt]
$i$ ($deg$)&74.86 $\pm$  0.35&71.41 $\pm$  0.34&52.71 $\pm$  1.17&48.17 
$\pm$  1.76&52.50&52.50&52.50&52.50\\[7pt]
$D_{P1}$ ($arcsec$)&0.437 $\pm$ 0.013&0.462 $\pm$ 0.014&0.427 $\pm$ 
0.004&0.456 $\pm$ 0.010&0.410 $\pm$ 0.014&0.445 $\pm$ 0.012&0.413 $\pm$ 
0.010&0.450 $\pm$ 0.019\\[7pt]
$D_{P2}$ ($arcsec$)&0.07989 $\pm$ 0.00005&0.07894 $\pm$ 0.00015&0.07231 
$\pm$ 0.00034&0.07156 $\pm$ 0.00042&0.07123 $\pm$ 0.00060&0.05108 $\pm$ 
0.00260&0.07157 $\pm$ 0.00007&0.07097 $\pm$ 0.00372\\[7pt]
$e_{max}$&0.133&0.177&0.198&0.177&0.231&0.309&0.151&0.218\\[7pt]
$\chi_\nu^2$& 4.40& 4.34& 7.22& 5.79&12.53& 9.48& 6.71& 6.50\\[7pt]
\enddata
\end{deluxetable}
\end{center}

\begin{center}
\begin{deluxetable}{lcccccccc}
\tablecolumns{9}
\tablewidth{0pt}
\rotate
\tabletypesize{\tiny}
\tablecaption{Parameter values for models fitting Data Set 2
\label{t.results2}}
\tablehead{{Model} & \colhead{9} & \colhead{10} & \colhead{11} &
\colhead{12} & \colhead{13} & \colhead{14} & \colhead{15} & 
\colhead{16}\\[7pt]
{Bulge} & \colhead{Weak} & \colhead{Weak} & \colhead{Strong} &
\colhead{Strong} & \colhead{Weak} & \colhead{Weak} & \colhead{Strong} & 
\colhead{Strong} \\[7pt]
{$F(e)$} & \colhead{Rayleigh} & \colhead{Gauss} & 
\colhead{Rayleigh} &
\colhead{Gauss} & \colhead{Rayleigh} & \colhead{Gauss} & 
\colhead{Rayleigh} & \colhead{Gauss} \\[7pt]
{Inclination} & \colhead{Free} & \colhead{Free} & 
\colhead{Free} &
\colhead{Free} & \colhead{Fixed} & \colhead{Fixed} & 
\colhead{Fixed} & \colhead{Fixed}}
\startdata
$M_{BH}$ ($\times 10^7 \msun$)&4.95 $\pm$ 0.08&4.67 $\pm$ 0.06&6.07 $\pm$ 
0.08&6.84 $\pm$ 0.06&5.35 $\pm$ 0.07&5.55 $\pm$ 0.07&5.73 $\pm$ 0.07&5.74 
$\pm$ 0.06\\[7pt]
$M_d$ ($\times 10^7 \msun$)&1.06 $\pm$ 0.03&1.39 $\pm$ 0.03&1.51 $\pm$ 
0.04&2.18 $\pm$ 0.05&1.57 $\pm$ 0.04&1.91 $\pm$ 0.05&1.96 $\pm$ 0.04&1.99 
$\pm$ 0.05\\[7pt]
$\Omega$ ($\kmspc$)&36.8 $\pm$  5.7&29.4 $\pm$  5.8&42.9 $\pm$ 11.9&46.7 
$\pm$  6.0&37.5 $\pm$  3.2&31.4 $\pm$  5.7&37.8 $\pm$  5.0&35.3 $\pm$  
3.0\\[7pt]
$\sigma_e$&0.1710 $\pm$ 0.0034&0.1940 $\pm$ 0.0038&0.1734 $\pm$ 
0.0051&0.2386 $\pm$ 0.0066&0.1818 $\pm$ 0.0036&0.2460 $\pm$ 0.0041&0.1241 
$\pm$ 0.0056&0.1926 $\pm$ 0.0060\\[7pt]
$\sigma_{\omega}$ ($rad$)&0.758 $\pm$ 0.035&0.721 $\pm$ 0.039&0.837 $\pm$ 
0.073&0.692 $\pm$ 0.030&0.886 $\pm$ 0.040&0.799 $\pm$ 0.029&1.070 $\pm$ 
0.038&0.779 $\pm$ 0.061\\[7pt]
$\sigma_a$ ($arcsec$)&0.0017 $\pm$ 0.0024&0.0030 $\pm$ 0.0012&0.0016 
$\pm$ 0.0027&0.0043 $\pm$ 0.0009&0.0042 $\pm$ 0.0013&0.0122 $\pm$ 
0.0015&0.0023 $\pm$ 0.0006&0.0030 $\pm$ 0.0049\\[7pt]
$\Delta$ ($arcsec$)&0.15 $\pm$ 0.09&0.28 $\pm$ 0.04&0.16 $\pm$ 0.03&0.07 
$\pm$ 0.07&0.08 $\pm$ 0.07&0.12 $\pm$ 0.13&0.18 $\pm$ 0.05&0.12 $\pm$ 
0.08\\[7pt]
$C$&0.431 $\pm$ 0.061&0.614 $\pm$ 0.035&0.587 $\pm$ 0.041&0.417 $\pm$ 
0.054&0.471 $\pm$ 0.085&0.462 $\pm$ 0.058&0.306 $\pm$ 0.053&0.509 $\pm$ 
0.158\\[7pt]
$R_d$ ($arcsec$)&3.00&3.00&3.00&3.00&3.79 $\pm$ 0.19&3.86 $\pm$ 0.15&4.16 
$\pm$ 0.10&3.33 $\pm$ 0.22\\[7pt]
$i$ ($deg$)&72.04 $\pm$  0.21&68.21 $\pm$  0.24&51.53 $\pm$  0.34&43.16 
$\pm$  0.35&52.50&52.50&52.50&52.50\\[7pt]
$D_{P1}$ ($arcsec$)&0.465 $\pm$ 0.010&0.427 $\pm$ 0.010&0.427 $\pm$ 
0.012&0.432 $\pm$ 0.005&0.413 $\pm$ 0.006&0.429 $\pm$ 0.008&0.452 $\pm$ 
0.005&0.467 $\pm$ 0.007\\[7pt]
$D_{P2}$ ($arcsec$)&0.07884 $\pm$ 0.00030&0.06853 $\pm$ 0.00030&0.07316 
$\pm$ 0.00006&0.06674 $\pm$ 0.00128&0.07152 $\pm$ 0.00003&0.05102 $\pm$ 
0.00014&0.06702 $\pm$ 0.00010&0.06333 $\pm$ 0.00021\\[7pt]
$e_{max}$&0.197&0.218&0.182&0.197&0.262&0.294&0.148&0.168\\[7pt]
$\chi_\nu^2$& 3.32& 2.61& 3.62& 2.66& 3.53& 2.77& 2.95& 2.73\\[7pt]
\enddata
\end{deluxetable}
\end{center}

\begin{center}
\begin{deluxetable}{lcccccccc}
\tablecolumns{9}
\tablewidth{0pt}
\rotate
\tabletypesize{\tiny}
\tablecaption{Parameter values for models fitting Data Set 3
\label{t.results3}}
\tablehead{{Model} & \colhead{17} & \colhead{18} & \colhead{19} &
\colhead{20} & \colhead{21} & \colhead{22} & \colhead{23} & 
\colhead{24}\\[7pt]
{Bulge} & \colhead{Weak} & \colhead{Weak} & \colhead{Strong} &
\colhead{Strong} & \colhead{Weak} & \colhead{Weak} & \colhead{Strong} & 
\colhead{Strong} \\[7pt]
{$F(e)$} & \colhead{Rayleigh} & \colhead{Gauss} & 
\colhead{Rayleigh} &
\colhead{Gauss} & \colhead{Rayleigh} & \colhead{Gauss} & 
\colhead{Rayleigh} & \colhead{Gauss} \\[7pt]
{Inclination} & \colhead{Free} & \colhead{Free} & 
\colhead{Free} &
\colhead{Free} & \colhead{Fixed} & \colhead{Fixed} & 
\colhead{Fixed} & \colhead{Fixed}}
\startdata
$M_{BH}$ ($\times 10^7 \msun$)&6.87 $\pm$ 0.06&5.38 $\pm$ 0.05&5.33 $\pm$ 
0.04&5.82 $\pm$ 0.05&5.36 $\pm$ 0.01&4.98 $\pm$ 0.03&4.68 $\pm$ 0.04&4.24 
$\pm$ 0.03\\[7pt]
$M_d$ ($\times 10^7 \msun$)&1.40 $\pm$ 0.02&1.46 $\pm$ 0.03&0.83 $\pm$ 
0.01&0.78 $\pm$ 0.01&1.58 $\pm$ 0.01&1.61 $\pm$ 0.01&1.06 $\pm$ 0.02&1.04 
$\pm$ 0.01\\[7pt]
$\Omega$ ($\kmspc$)&55.6 $\pm$  2.5&33.6 $\pm$  3.1&27.1 $\pm$  1.3&38.2 
$\pm$  2.9&42.4 $\pm$  0.5&27.6 $\pm$  1.9&24.5 $\pm$  1.0&26.1 $\pm$  
1.0\\[7pt]
$\sigma_e$&0.2886 $\pm$ 0.0010&0.2923 $\pm$ 0.0013&0.1691 $\pm$ 
0.0012&0.3129 $\pm$ 0.0026&0.2040 $\pm$ 0.0003&0.2327 $\pm$ 0.0009&0.2634 
$\pm$ 0.0014&0.3597 $\pm$ 0.0018\\[7pt]
$\sigma_{\omega}$ ($rad$)&0.746 $\pm$ 0.007&0.784 $\pm$ 0.009&0.784 $\pm$ 
0.015&0.718 $\pm$ 0.015&0.823 $\pm$ 0.011&0.821 $\pm$ 0.010&1.008 $\pm$ 
0.009&0.554 $\pm$ 0.003\\[7pt]
$\sigma_a$ ($arcsec$)&0.0009 $\pm$ 0.0001&0.0022 $\pm$ 0.0001&0.0058 
$\pm$ 0.0009&0.0086 $\pm$ 0.0018&0.0005 $\pm$ 0.0003&0.0070 $\pm$ 
0.0009&0.0292 $\pm$ 0.0074&0.0374 $\pm$ 0.0046\\[7pt]
$\Delta$ ($arcsec$)&0.06 $\pm$ 0.10&0.25 $\pm$ 0.13&0.20 $\pm$ 0.05&0.12 
$\pm$ 0.05&0.04 $\pm$ 0.07&0.14 $\pm$ 0.02&0.26 $\pm$ 0.04&0.02 $\pm$ 
0.09\\[7pt]
$C$&0.384 $\pm$ 0.011&0.625 $\pm$ 0.006&0.593 $\pm$ 0.009&0.492 $\pm$ 
0.011&0.445 $\pm$ 0.026&0.467 $\pm$ 0.019&0.263 $\pm$ 0.029&0.466 $\pm$ 
0.032\\[7pt]
$R_d$ ($arcsec$)&3.00&3.00&3.00&3.00&4.20 $\pm$ 0.04&4.11 $\pm$ 0.05&3.77 
$\pm$ 0.04&2.98 $\pm$ 0.04\\[7pt]
$i$ ($deg$)&41.31 $\pm$  0.15&49.43 $\pm$  0.14&62.57 $\pm$  0.08&61.23 
$\pm$  0.14&52.50&52.50&52.50&52.50\\[7pt]
$D_{P1}$ ($arcsec$)&0.483 $\pm$ 0.002&0.456 $\pm$ 0.003&0.545 $\pm$ 
0.006&0.465 $\pm$ 0.007&0.435 $\pm$ 0.001&0.426 $\pm$ 0.004&0.679 $\pm$ 
0.007&0.775 $\pm$ 0.009\\[7pt]
$D_{P2}$ ($arcsec$)&0.07874 $\pm$ 0.00109&0.06569 $\pm$ 0.00084&0.07116 
$\pm$ 0.00028&0.06725 $\pm$ 0.00004&0.07100 $\pm$ 0.00093&0.05429 $\pm$ 
0.00071&0.05672 $\pm$ 0.00029&0.06831 $\pm$ 0.00033\\[7pt]
$e_{max}$&0.086&0.116&0.088&0.004&0.065&0.091&0.214&0.306\\[7pt]
$\chi_\nu^2$&10.56&10.98&15.55&18.42&10.32&11.33&15.90&15.21\\[7pt]
\enddata
\end{deluxetable}
\end{center}

\begin{figure}
\plotone{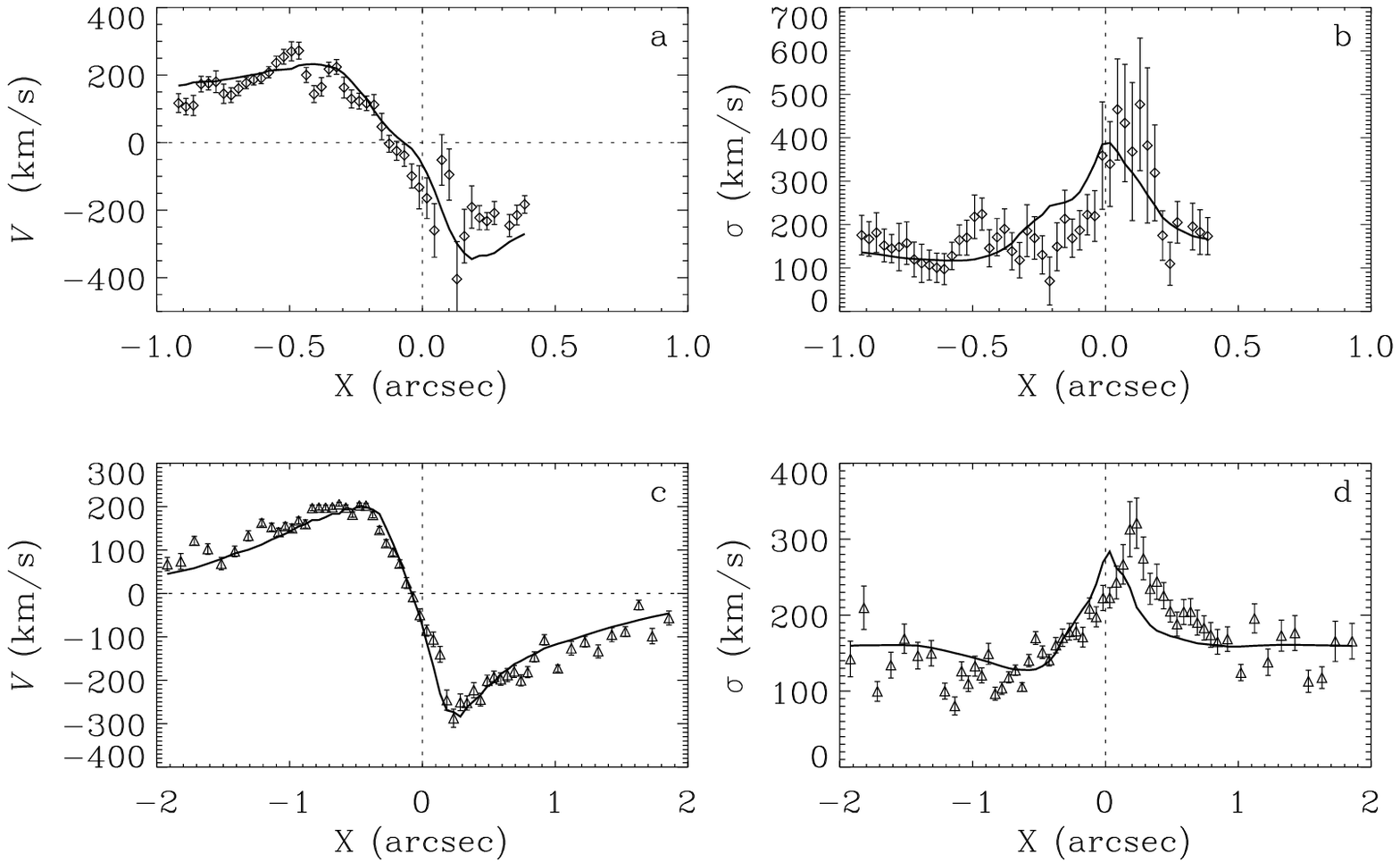}
\caption{\footnotesize
Solid lines show one-dimensional kinematic profiles for Model 2, which is
representative of models in Table \ref{t.results1}.  Shown are the (a)
rotation curve and (b) velocity dispersion profile at FOC resolution, and
the (c) rotation curve and (d) velocity dispersion profile at STIS
resolution.  FOC data from Statler et al. (1999) are shown as diamonds and
STIS data from B01 are shown as triangles.  The UV peak is at the origin. 
The model reproduces the asymmetric rotation amplitudes, the offset
zero-velocity crossing, and the low velocity dispersion at $\sim 0\farcs
5$.  The location of the peak in velocity dispersion is problematic, and
is discussed in Section \ref{s.bhresults} and Section \ref{sec:CHAP5}. 
}
\label{f.1dkinbest1}
\end{figure}

\begin{figure}
\plotone{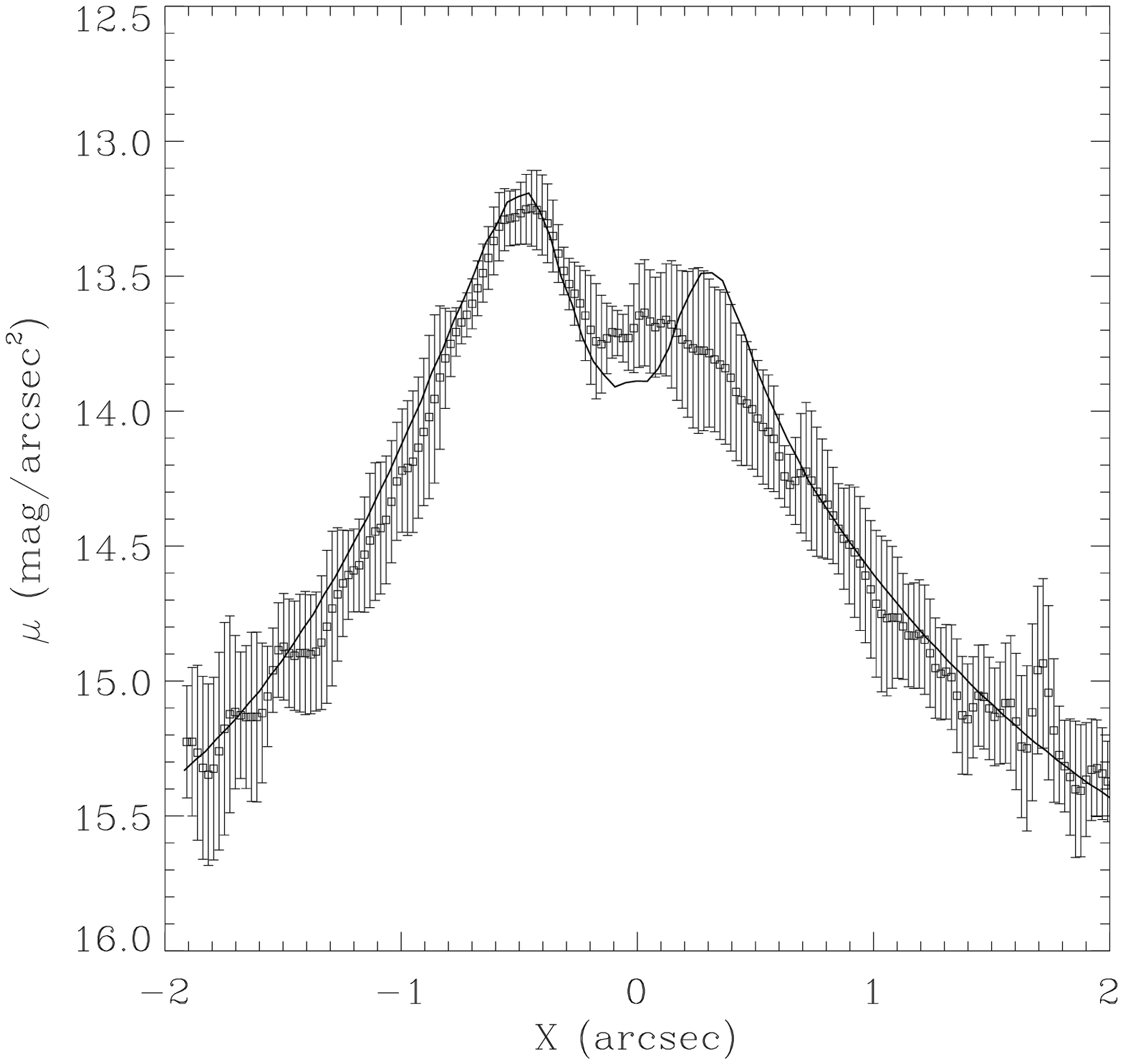}
\caption{\footnotesize
The solid line shows the one-dimensional photometric profile for Model 2
in Table \ref{t.results1}.  Squares show $I$ band WFPC2/HST data (L98),
averaged over a slit of width $0\farcs 353$ and pixel scale $l=0\farcs
0456$, at position angle {\it PA}$=52.5\arcdeg$ (as in KB99); we refer to 
this as 1WFPC2 data.  The UV peak is at the origin.  The shape of the
brightness profile near P1 and outside $0\farcs6$ is reproduced by the
model. 
}
\label{f.1dphotbest1}
\end{figure}

Figures \ref{f.1dkinbest1} and \ref{f.1dphotbest1} show one-dimensional
kinematic and photometric profiles for Model 2 in Table \ref{t.results1}. 
The curves in panels (a) and (b) of Figure \ref{f.1dkinbest1} show the model
rotation curve and velocity dispersion profile at FOC resolution; Diamonds
show FOC data for comparison.  The rotation curve and dispersion profile
at STIS resolution are given in panels (c) and (d) of Figure
\ref{f.1dkinbest1};  STIS data are shown as triangles.  Modeled
one-dimensional HST photometry is shown as the curve in figure
\ref{f.1dphotbest1};  squares show the 1WFPC2 data points described in 
Section \ref{s.datasample}. 

Many of the important features in the observed profiles are reproduced by
the model.  These features include the asymmetric rotation amplitudes in
both the FOC and STIS profiles, the offset zero-velocity crossing (ZVC),
the low velocity dispersion at $\sim -0\farcs5$, and the shape of 
the brightness profile near P1 and outside $0\farcs6$.

The detailed shape of the FOC rotation curve near $v = 0 \kms$ is not
exactly reproduced by the model;  but this part of the profile could be
improved by adding a small amount of rotation to the bulge.  More
conspicuously, the position of maximum velocity dispersion is not
reproduced, especially in the STIS data.  This is a ubiquitous property of
all of the fits.  It is not clear whether the problem lies with the models
or with the data.  We defer discussion of this issue to Section
\ref{s.bhresults} and Section \ref{sec:CHAP5}, and focus here on the
amplitude of the dispersion peak. 

\begin{figure}
\plotone{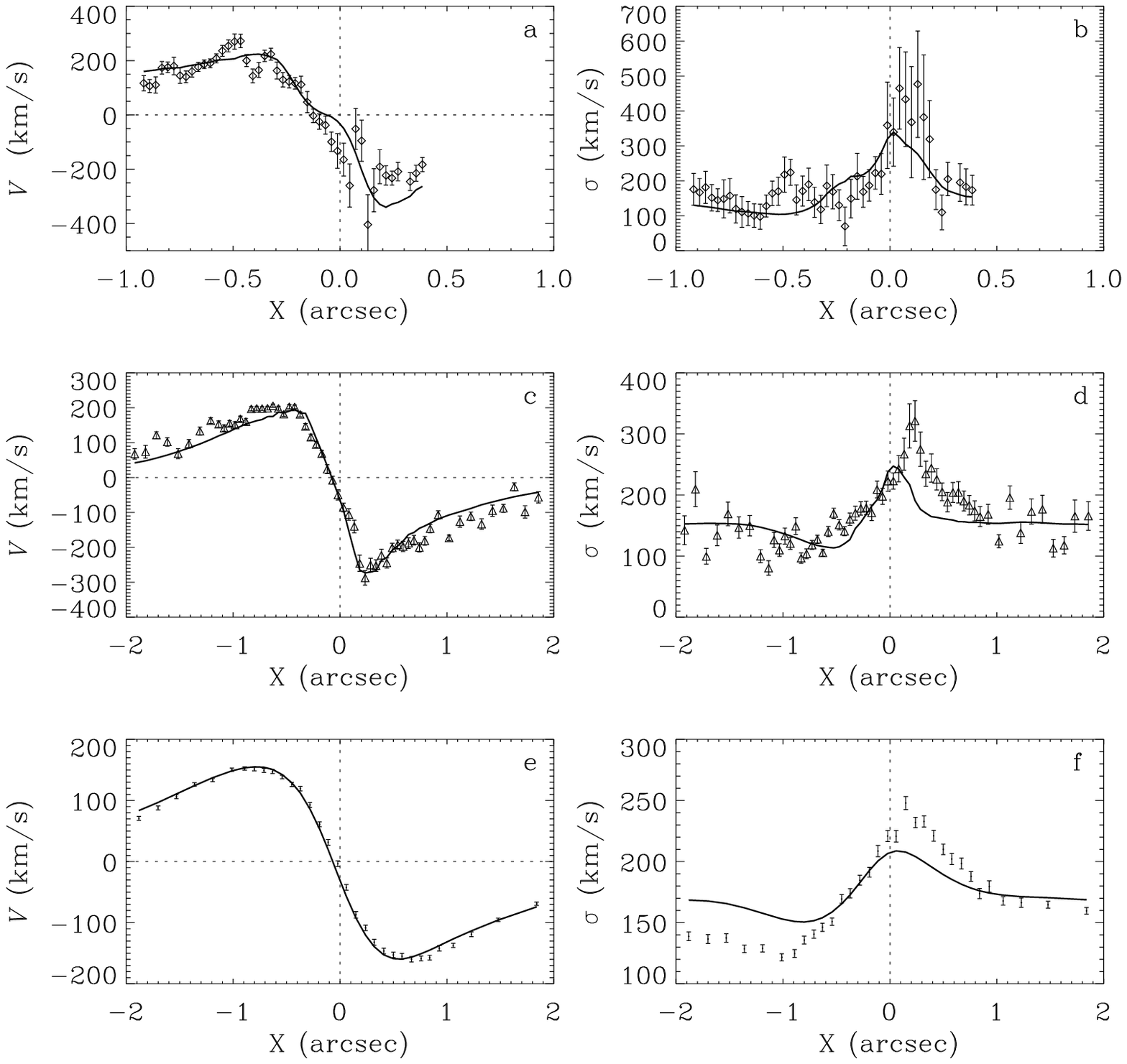}
\caption{\footnotesize
Solid lines show one-dimensional kinematic profiles for Model 10, a
representative model from Table \ref{t.results2}.  FOC and STIS velocity
profiles are shown in Panels (a) through (d), as in Figure
\ref{f.1dkinbest1}.  Panels (e) and (f) show the rotation curve and
velocity dispersion at SIS resolution, respectively; SIS data from KB99 are
shown as error bars.  Model 10 is similar to Model 2, since parameters
shift by $30\%$ at most when SIS and OASIS kinematics are included in the 
fit. }
\label{f.1dkinbest2}
\end{figure}

\begin{figure}
\plotone{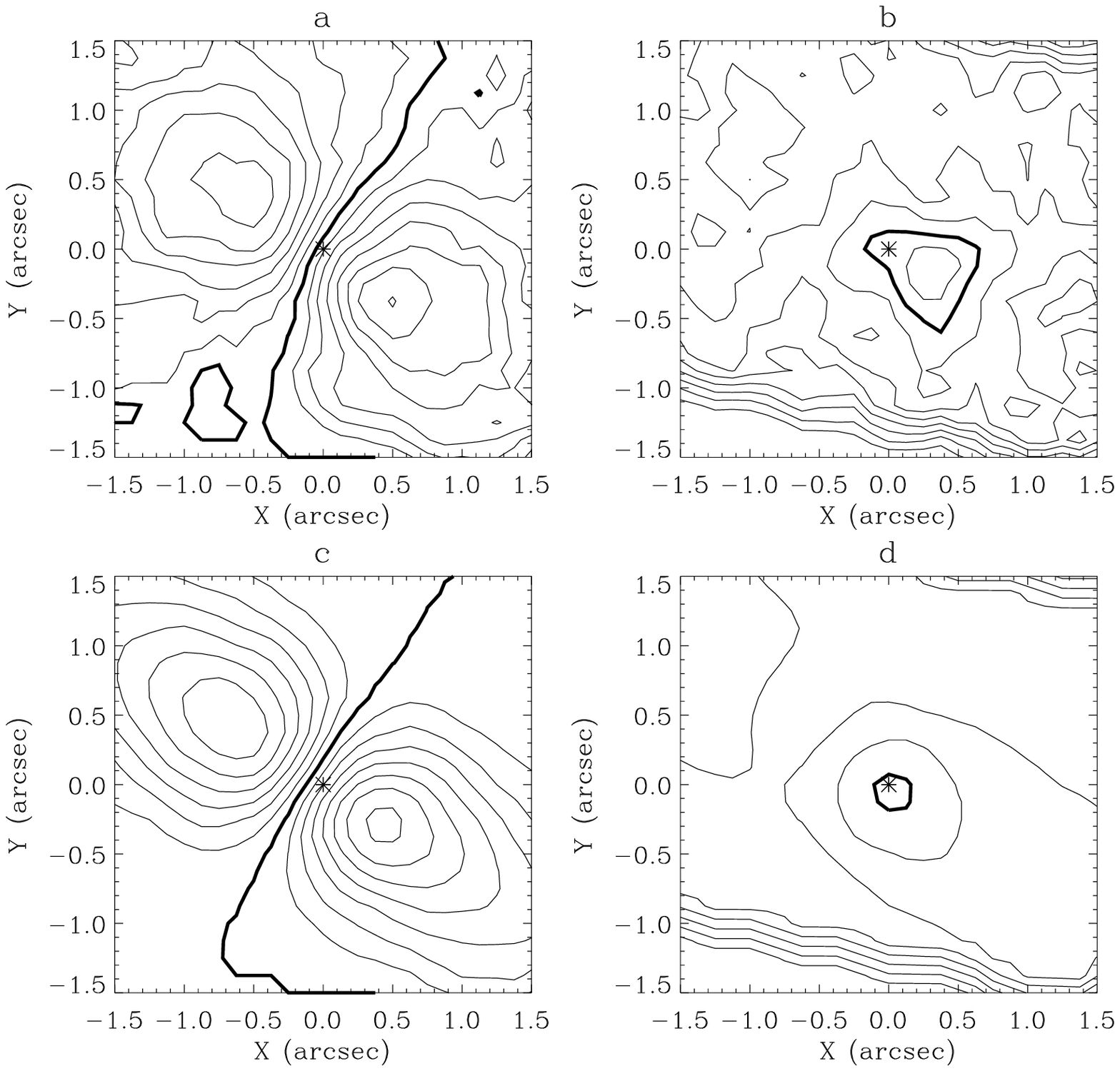}
\caption{\footnotesize
Two-dimensional kinematic profiles for Model 10 in Table \ref{t.results2}. 
Shown are the (a) mean-velocity field and (b) velocity dispersion field
from OASIS (B01), and the (c) mean-velocity field and (d) velocity
dispersion field of the model.  Mean-velocity contours run from -250\kms\
to 250\kms\ in steps of 25\kms.  Velocity dispersion contours run from
0\kms to 300\kms in steps of 25\kms.  The thick line shows the zero
isovelocity contour and the 200\kms\ isovelocity dispersion.  The UV peak
is labeled with an asterisk.  The kinematic axis is at {\it
PA}$_K=56.4\arcdeg$ and {\it PA}$_K=55.3\arcdeg$ in the data and model,
respectively; this validates our choice of equating the {\it PA} of the
line-of-nodes with {\it PA}$_K$ (Section \ref{s.assume}). 
}
\label{f.2dkinbest2}
\end{figure}

Figure \ref{f.1dkinbest2} shows one-dimensional kinematic profiles for
model 10 in Table \ref{t.results2}.  The photometric profile is very
similar to Figure \ref{f.1dphotbest1}.  Figure \ref{f.1dkinbest2} includes
SIS rotation and dispersion profiles in panels (e) and (f).  Panels (a)
and (b) of Figure \ref{f.2dkinbest2} show two-dimensional OASIS
mean-velocity and velocity dispersion fields, while panels (c) and (d)
show the corresponding model kinematic profiles.  One-dimensional model
FOC and STIS kinematic profiles for Model 10 are similar to those found
for Model 2.  Parameters shift by at most $30\%$ when SIS and OASIS 
kinematics are
added in the fitting routine.  The SIS rotation curve is well reproduced
by the model, but the dispersion profile is not as well fit.  The
kinematic major axis is {\it PA}$_K = 55.3\arcdeg$ for Model 10, which is 
close to
the measured value of $56.4\arcdeg$ (B01).  The assumption that the
line-of-nodes is at {\it PA}$_n = 56.4\arcdeg$ appears valid (see Section
\ref{s.assume}). 

\begin{figure}
\plotone{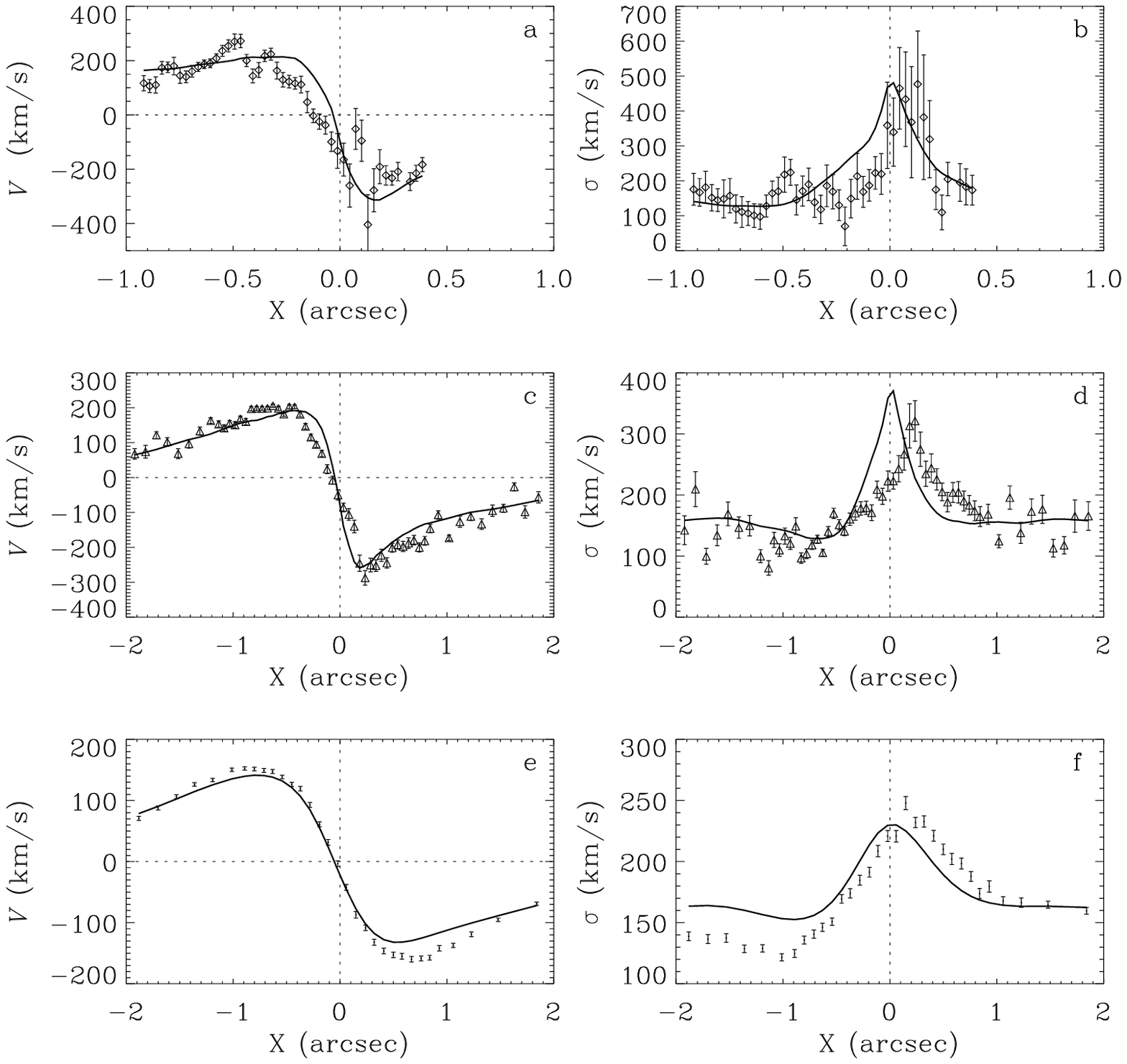}
\caption{\footnotesize
Solid lines show one-dimensional kinematic profiles for Model 17, a
representative model from Table \ref{t.results3}.  Panels (a) through (f)
show FOC, STIS, and SIS profiles, as in Figure \ref{f.1dkinbest2}.  The
quality of the fit diminishes when two-dimensional photometry is added in
the fit;  compare this plot with Figures \ref{f.1dkinbest1} (Model 2) and
\ref{f.1dkinbest2} (Model 10).  Models with $i\sim 50 \arcdeg$, like Model 
17, are better able to fit the amplitude of the dispersion spike. 
}
\label{f.1dkinbest3}
\end{figure}

\begin{figure}
\plotone{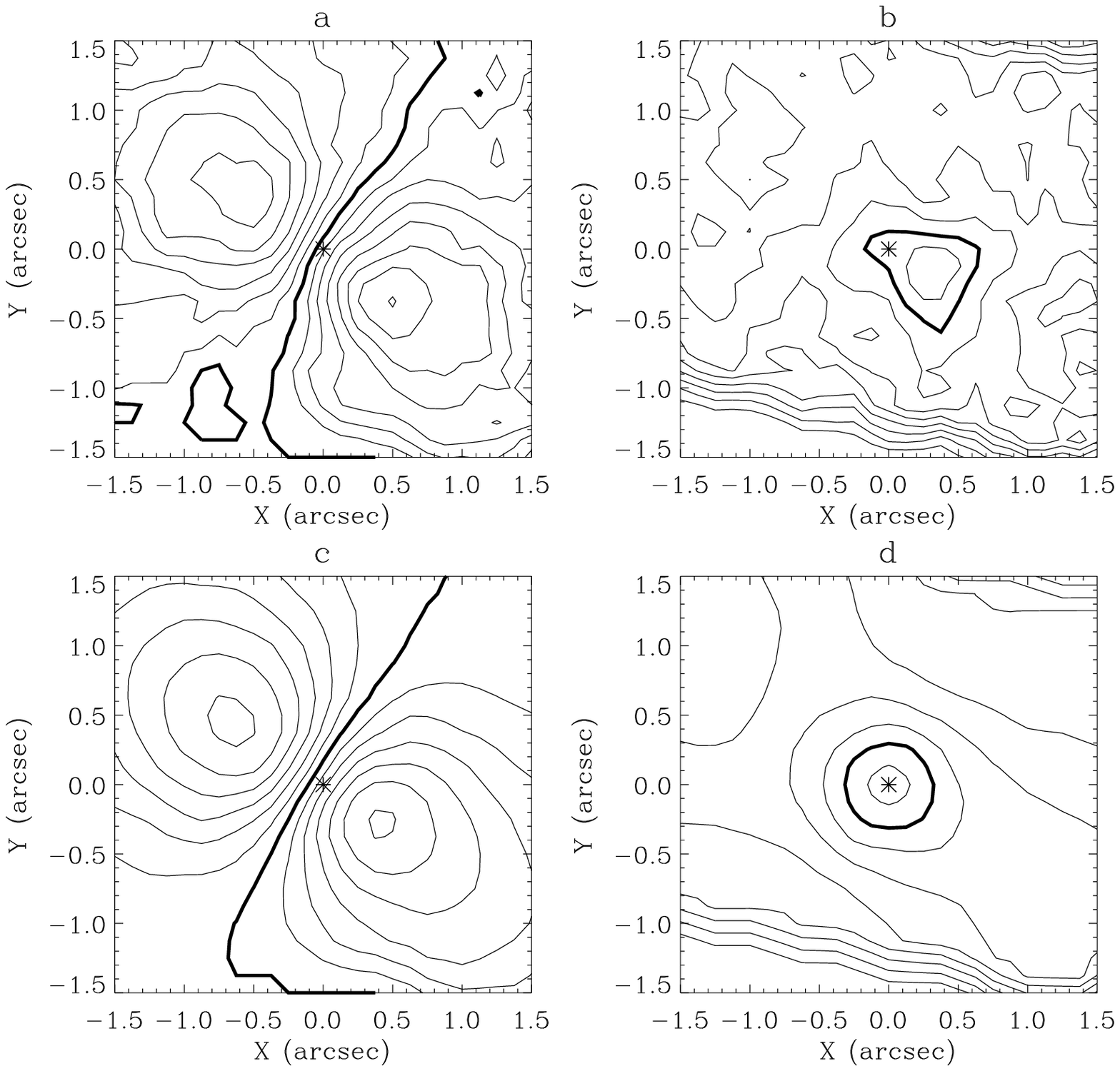}
\caption{\footnotesize
Two-dimensional kinematic profiles for Model 17 in Table \ref{t.results3}. 
Shown are the (a) mean-velocity field and (b) velocity dispersion field
from OASIS, and the (c) mean-velocity field and (d) velocity dispersion
field for the model.  Mean-velocity contours run from -250\kms\ to 250\kms\
in steps of 25\kms.  Velocity dispersion contours run from 0\kms\ to
300\kms\ in steps of 25\kms.  The thick line shows the zero isovelocity
contour and the 200\kms\ isovelocity dispersion.  The UV peak is labeled
with an asterisk.  Models with $i \sim 50 \arcdeg$ provide a better match 
to the OASIS velocity map;  compare with Figure \ref{f.2dkinbest2}.
} 
\label{f.2dkinbest3}
\end{figure}

\begin{figure}[htb]
\plotone{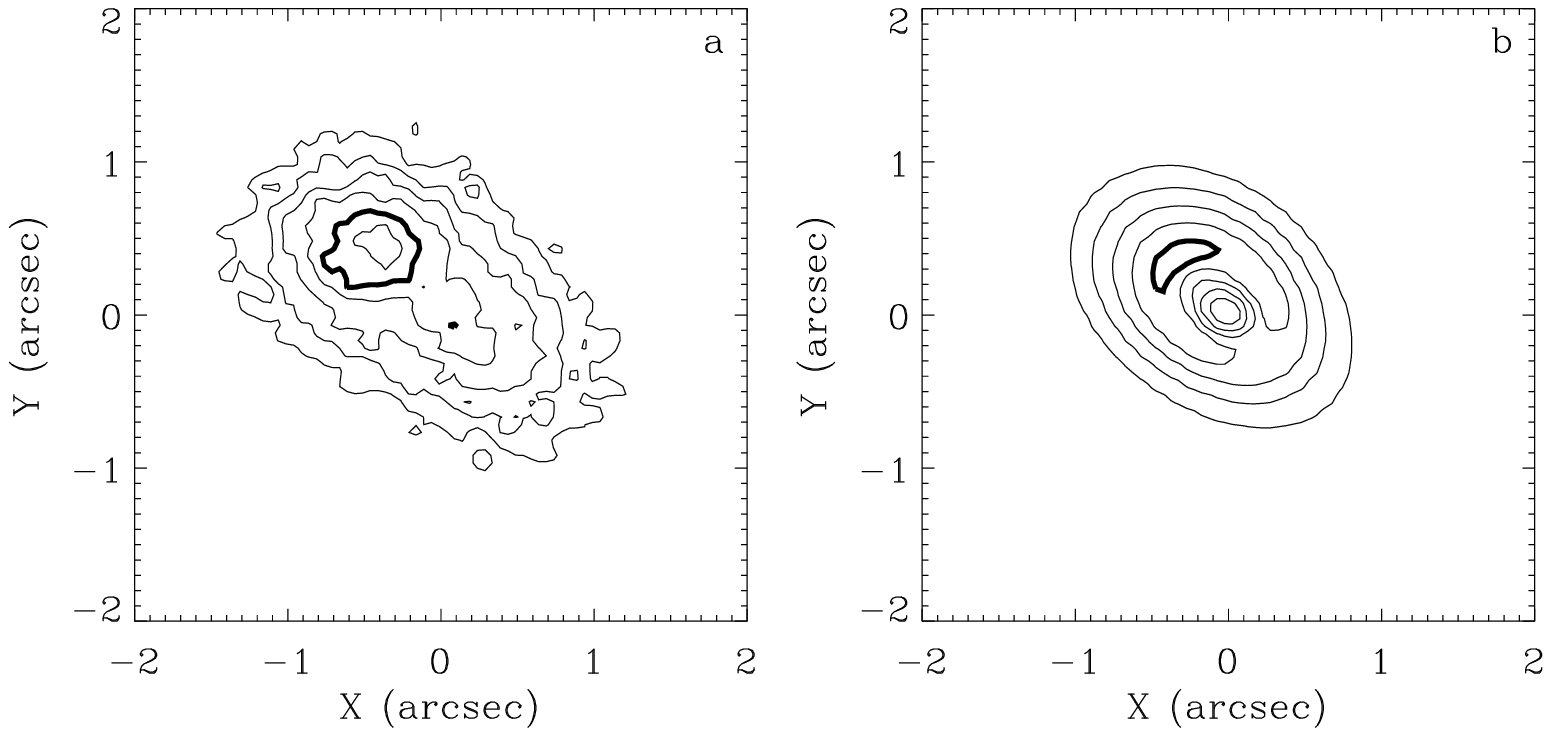}
\caption{\footnotesize
Two-dimensional photometric profile for Model 17 in Table
\ref{t.results3}.  Panel (a) shows 2WFPC2 data, which is the $I$ band
WFPC2/HST data from L98, binned on an $80\times80$ grid with spacing
$0\farcs05$; panel (b) shows the corresponding model surface brightness. 
Contours run from 14\magarc\ to 12\magarc\ in steps of 0.25\magarc.  The
thick line shows the 13.0\magarc\ contour.  Two-dimensional models possess
crescent-shaped P1 distributions;  see Section \ref{sec:CHAP5} for a
discussion.  Model 17 has a weak bulge, so the central surface 
brightness is weak.
}
\label{f.2dphotbest3}
\end{figure}

Figures \ref{f.1dkinbest3} and \ref{f.2dkinbest3} show one and
two-dimensional kinematic profiles for Model 17 in Table \ref{t.results3}. 
Figure \ref{f.2dphotbest3} shows two-dimensional photometry from 2WFPC2; 
panel (a) shows the data, while panel (b) shows the corresponding plot for
the model.  Model 17 does not fit the one-dimensional kinematic profiles
as well as Model 10.  Replacing 1WFPC2 with 2WFPC2 when moving from Data
Set 2 to Data Set 3 can cause disk parameters to shift by more than
$100\%$ in some cases.  However, the range of BH masses is not
significantly altered by the changes.  The mean-velocity map for Model 17
fits the OASIS data well; the contours are more circular than those found
for Model 10.  The kinematic axis is at $56.9\arcdeg$, similar to its
value in the data. 

Figure \ref{f.2dphotbest3} shows that the surface brightness distribution
of Model 17 has a prominent P1 structure, but that it has a crescent
shape.  Crescent-shaped brightness distributions are found in our
models, and are probably a result of limiting the model to two-dimensions. 
A discussion on this point is given in Section
\ref{sec:CHAP5}. 

The kinematic profiles in Figures \ref{f.1dkinbest1} through
\ref{f.2dkinbest3} suggest that the $\sim 70\arcdeg$ inclination of Models
2 and 10 is too large.  Models with $i \sim 50\arcdeg$, like Model 17, are
better able to fit the amplitude of the dispersion spike (Figure
\ref{f.1dkinbest3}) and the OASIS velocity map (Figure
\ref{f.2dkinbest3}). 

\begin{figure}
\plotone{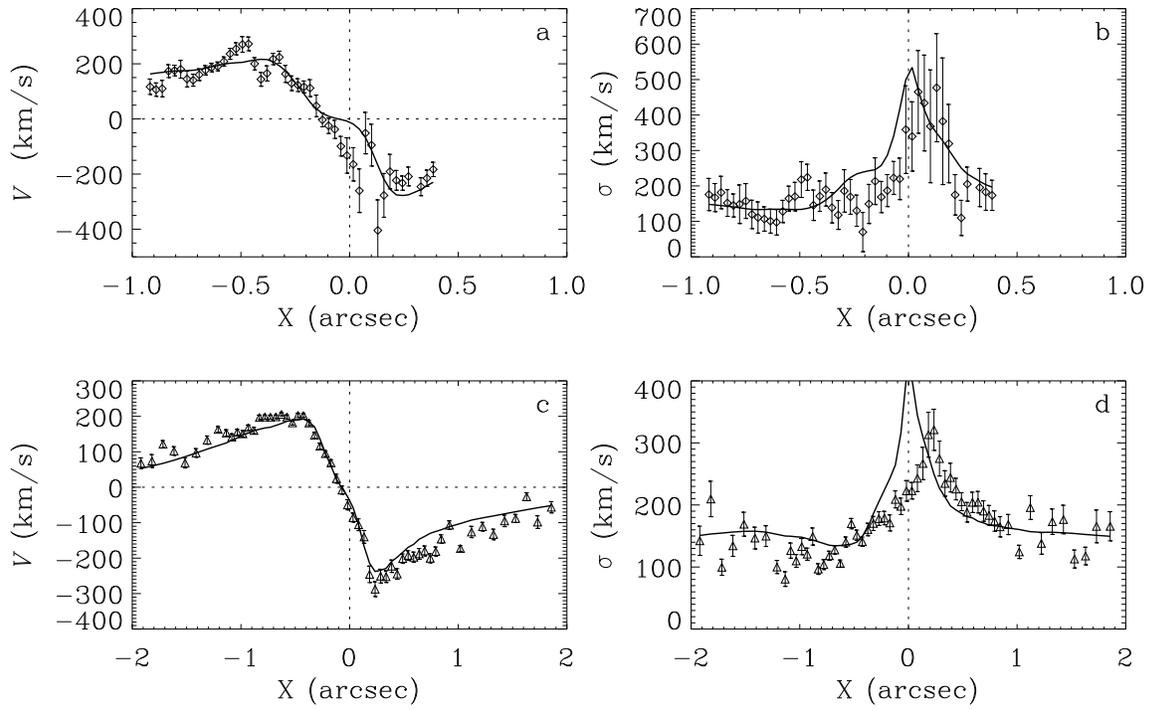}
\caption{\footnotesize
Solid lines show one-dimensional kinematic profiles for Model 4 in Table
\ref{t.results1}, which has a strong bulge; Models 10 (Figure
\ref{f.1dkinbest2}) and 17 (Figure \ref{f.1dkinbest3}) have a weak bulge. 
A stronger bulge cusp can diminish over-rotation near the origin in models
with $i\sim50\arcdeg$.  The bulge is too strong here, but the effect is
clearly demonstrated in the FOC rotation curve in Panel (a). 
}
\label{f.1dkinalt12}
\end{figure}

The rotation curve for models with $i \sim 50 \arcdeg$ and a weak bulge
typically over-rotates inside $0\farcs4$, as seen in Figure
\ref{f.1dkinbest3}.  A stronger bulge cusp may improve the fit to
the inner rotation curve for low-inclination models.  Figure
\ref{f.1dkinalt12} shows kinematic profiles for Model 4 in Table
\ref{t.results1}, which includes the strong bulge model.  The strong bulge
model is too strong in this case, as can be seen from the
nearly-flat FOC rotation curve near $x=0\arcsec$ (panel a), but it is
clear that a stronger inner bulge can lessen over-rotation in the central
regions.  A stronger bulge can also improve the
fit to the surface photometry near the UV peak and P2.  The strength of
the central dip between P1 and P2 (Figures \ref{f.1dphotbest1} and
\ref{f.2dphotbest3}) increases when the inclination is reduced.  A
stronger bulge cusp can fill in the missing light in the
hole, but at the cost of flattening the rotation curve near the origin. 

\subsection{The Supermassive Black Hole in M31}
\label{s.bhresults}

Table \ref{t.weightavg} gives the weighted averages and total
uncertainties of parameter values in each of the three Tables
(\ref{t.results1}, \ref{t.results2}, and \ref{t.results3}) taken
separately, and altogether as one combined grid of models.  The total
uncertainty is given by the quadrature sum of the statistical and
systematic uncertainites.  The statistical uncertainty is given by the
weighted average of the statistical errors in each model.  The systematic
uncertainty is given by the weighted standard deviation of the best-fit
values in each table.  Weighted averages and uncertainties for $R_d$ and
$i$ include only those models for which the parameter was free in the
fitting.  A mean and standard deviation is given for $e_{max}$. 

\begin{deluxetable}{lcccc}
\tablecolumns{5}
\tablewidth{0pt}
\tabletypesize{\scriptsize}
\tablecaption{Weighted averages and total uncertainties
\label{t.weightavg}}
\tablehead{
{Parameter} & \colhead{Table \ref{t.results1}} & \colhead{Table 
\ref{t.results2}} & \colhead{Table \ref{t.results3}} & \colhead{All Tables}}
\startdata
$M_{BH}$ ($\times 10^7 \msun$)&6.04 (0.24)&5.62 (0.66)&5.24 (0.43)&5.29 
(0.49)\\[7pt]
$M_d$ ($\times 10^7 \msun$)&1.54 (0.26)&1.57 (0.38)&1.34 (0.33)&1.36 
(0.34)\\[7pt]
$\Omega$ ($\kmspc$)&40.9 (6.3)&36.5 (4.2)&36.7 (8.3)&36.7 (8.1)\\[7pt]
$\sigma_e$&0.2047 (0.0261)&0.1894 (0.0323)&0.2220 (0.0383)&0.2208 
(0.0384)\\[7pt]
$\sigma_{\omega}$ ($rad$)&0.717 (0.086)&0.806 (0.115)&0.666 (0.142)&0.671 
(0.142)\\[7pt]
$\sigma_a$ ($arcsec$)&0.0019 (0.0013)&0.0036 (0.0025)&0.0014 
(0.0012)&0.0016 (0.0014)\\[7pt]
$\Delta$ ($arcsec$)&0.12 (0.05)&0.17 (0.07)&0.15 (0.06)&0.15 (0.06)\\[7pt]
$C$&0.515 (0.055)&0.505 (0.109)&0.554 (0.094)&0.551 (0.094)\\[7pt]
$R_d$ ($arcsec$)&3.59 (0.24)&3.96 (0.27)&3.69 (0.51)&3.71 (0.50)\\[7pt]
$i$ ($deg$)&71.81 (5.44)&63.51 (10.80)&56.89 (8.10)&58.56 (9.24)\\[7pt]
$D_{P1}$ ($arcsec$)&0.431 (0.014)&0.438 (0.017)&0.453 (0.050)&0.450 
(0.046)\\[7pt]
$D_{P2}$ ($arcsec$)&0.07663 (0.00401)&0.07063 (0.00424)&0.06711 
(0.00181)&0.07125 (0.00503)\\[7pt]
$e_{max}$&0.199 (0.055)&0.208 (0.049)&0.121 (0.095)&0.176 (0.077)\\[7pt]
\enddata
\tablecomments{Weighted averages and total uncertainties
(parentheses) for fitted parameters in Tables \ref{t.results1},
\ref{t.results2}, and \ref{t.results3} separately, and taken all together. 
The total uncertainty is given by the quadrature sum of the statistical
and systematic uncertainties, as described in Section \ref{s.bhresults}. 
Weighted averages and uncertainties for $R_d$ and $i$ include only models
for which that parameter was free.  A mean and standard deviation is found
for $e_{max}$.}
\end{deluxetable}

We take the averages and uncertainties computed from Table
\ref{t.results2} (Data Set 2) as the statement of our best results.  Data
Set 3 is dominated by photometric data outside $1 \arcsec$, which is
somewhat poor in quality and de-emphasizes the disk asymmetry.  This also
applies to the results for the full grid of models (the fourth column of
Table \ref{t.weightavg}), since results from Data Set 3 dominate in
weighted averages due to their small errors.  Results from Data Set 2 are
also consistent, to roughly $1 \sigma$, with results from the other two
Data Sets. 

The mass of the BH in M31 is thus $5.62 \pm 0.66 \times 10^7 \msun$. 
Other authors find $M_{BH}$ values of $0.1-1 \times 10^7 \msun$ (Dressler
1984), $3-7 \times 10^7 \msun$ (Dressler \& Richstone 1988), $0.3-10
\times 10^7 \msun$ (Kormendy 1988), $4-5 \times 10^7 \msun$ (Richstone,
Bower, \& Dressler 1990), $7 \times 10^7 \msun$ (Bacon et al. 1994), $7.5
\times 10^7 \msun$ (T95), $7-10 \times 10^7 \msun$ (Emsellem \& Combes
1997), $3.3 \pm 1.5 \times 10^7 \msun$ (KB99), $3.5-8.5 \times 10^7 \msun$
(B01), and $10.2 \times 10^7 \msun$ (PT03). 

Results from Table \ref{t.weightavg} suggest that the BH is located in the
UV peak.  The parameter $D_{P2}$ gives the BH-P2 separation along the disk
major axis.  We find $D_{P2} = 0 \farcs 071 \pm 0 \farcs 004$, which is
close to the $0\farcs076$ P2-UV peak separation measured by B01.  Also,
the measured P1-UV peak separation of $0\farcs 44$ (B01) is consistent
with the $0 \farcs 438 \pm 0 \farcs 017$ value for the P1-BH separation,
$D_{P1}$.  The UV peak does not lie along the major axis ({\it PA}$_d$). 
There is a $\sim 0\farcs 02$ perpendicular offset between the {\it PA}$_d$
line and the UV peak.  When projected onto the P1-P2 axis, the P1-UV peak
and UV peak-P2 separations are $0\farcs 439$ and $0\farcs 074$,
respectively, which are consistent with $D_{P1}$ and $D_{P2}$.  The UV
peak has a $\sim 0\farcs 2$ half-power width, so the perpendicular offset
is negligible. 

We find that the location of the spike in velocity dispersion in the
models is always close to that of the BH.  Physically, this is expected,
since the bulge dispersion must peak near the BH if the latter dominates
the gravity, and disk material orbiting close to the BH will produce the
same effect.  Since the BH is in the UV peak and not near P2, where the
spike is found in the data, our models are not able to reproduce the
offset location of the dispersion spike.  Three possible explanations for
this inconsistency include:  first, that there is a problem with the
positional registration of the data;  second, that the models are correct
in essence but missing an essential component, such as retrograde orbits;
third, that the basic assumptions of the model incorrectly describe
the double nucleus in M31.  Further discussion will be
given on these points in Section \ref{sec:CHAP5}. 

\subsection{Disk Properties}
\label{s.diskprops}

The mass of the eccentric disk in M31 is $M_d = 1.57 \pm 0.38 \times 10^7 
\msun$ (within $\sim 14$ pc).  T95 finds $M_d=1.2 \times 10^7 \msun$ (within
$5.5$ pc) for his simple model consisting of three Keplerian ringlets. 
B01 find $M_d=1.7 \times 10^7 \msun$ (within 10 pc) in their N-body
simulation of an $m=1$ mode in a cold disk.  SS02 find $M_d=1.4 \times
10^7 \msun$ for a disk constructed using a Schwarzschild-type method.  The
recent photometric decomposition in P02 gives $2.1 \times 10^7 \msun$ for 
the sum of P1 and P2. 

The disk rapidly precesses at speed $\Omega = 36.5 \pm 4.2 \kmspc$; 
corotation is at $r \sim 1 \farcs 52$ for this $\Omega$.  Model disks in
papers from other authors have precession rates of $3 \kmspc$ (B01), $16
\kmspc$ (SS02), and $\sim 17 \kmspc$ (for $\epsilon
\sim 0.28$ in N-body simulations of lopsided modes in annular disks; 
Jacobs \& Sellwood 2001);  Sambhus \& Sridhar (2000) find $\Omega = 34 \pm
8 \kmspc$ and $\Omega = 20 \pm 12 \kmspc$ using a variant on the Tremaine
\& Weinberg (1984) method for two different fits to the bulge. 

\begin{figure}[tb]
\plotone{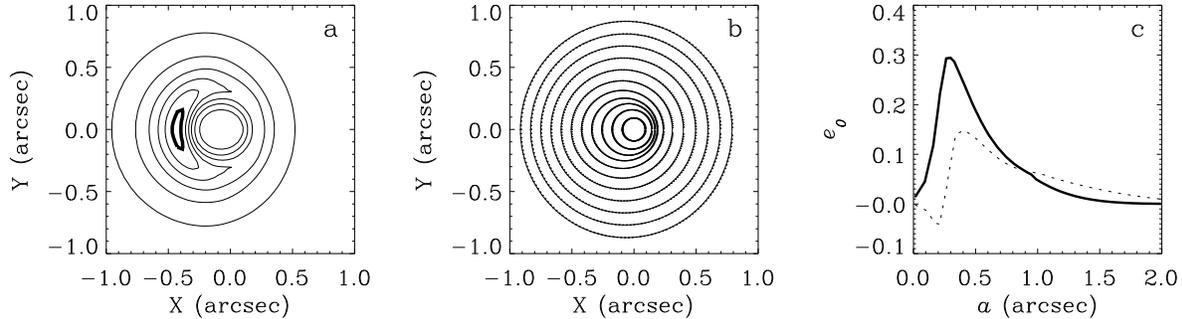}
\caption{\footnotesize
(a) Disk-only density contours for Model 14 in Table
\ref{t.results2}, which has $M_{BH} = 5.55 \times 10^7 \msun$, similar to
our overall best-fit value of $5.62 \pm 0.66 \times 10^7 \msun$.  Contours
are at 0.2, 0.35, ..., 0.95 of the maximum density.  The $95\%$ contour is
labeled with a thick line.  The central point mass is at (0,0), near the
point of minimum density.  (b) Uniformly precessing periodic orbits in the
total potential.  The radial variation of eccentricity is a consequence of
disk self-gravity.  (c) The solid line shows the eccentricity of the
orbits in (b) plotted against the semimajor axis;  this is the function
$e_0(a)$ in Equations \ref{e.fofeg} and \ref{e.fofer}.  The dotted line
shows $e_0(a)$ for Model 16, which has backbone orbits that switch
apoapses to the anti-P1 side of the BH at low semimajor axis;  many of our
models share this behavior.}
\label{f.densorbeofa}
\end{figure}

Figure \ref{f.densorbeofa} shows a contour plot of the surface density
(panel a), the set of backbone orbits (panel b), and the function $e_0(a)$
(solid line in panel c) describing the orbit sequence for the disk of
Model 14 in Table \ref{t.results2}.  Model 14 provides a good example of
the properties exhibited by models fitting Data Set 2, and has a BH mass
close to the weighted average in Table \ref{t.weightavg}. 

The non-axisymmetric density distribution shown in panel (a) is typical of
that found in our fits.  The strong density minimum near the origin is
indicative of a narrow radial mass distribution ($\sigma_a = 0\farcs 012$)
and a large central hole ($\Delta = 0\farcs 12$). 

The shape of the backbone orbit sequence shown in panel (c) is similar for
models with Gaussian and Rayleigh $F(e)$s.  The sequence of orbits
follows a steep eccentricity gradient through the densest part of the disk
($a \sim 0\farcs 4$), but there is no tendency for the sequence to reverse
apoapses to the anti-P1 side of the disk following this gradient 
(making $e$ negative), as 
found in S99 and Salow \& Statler (2001).  Even though models fitting
M31 do not show an eccentricity sign reversal, such models do exist for
lower values of $\Omega$;  see Section \ref{sec:CHAP5} for a discussion. 
An eccentricity reversal is found in some models inside $a \sim
0\farcs15$, but the minimum eccentricity never dips below $e_{min}=-0.05$
(see the dotted line in Figure \ref{f.densorbeofa}{c} for an example). 
Backbone sequences similar to ours are found for models in B01, SS02, and
PT03, except for the small eccentricity reversal at low semi-major axis in
some models.  The peak eccentricity is small ($e_{max}=0.294$);  other
authors find $e_{max}$ values of $\sim 0.7$ (B01), $\sim 0.7$ (SS02), and
$\sim 0.6$ (PT03).  Disk asymmetry, and thus $e_{max}$, is most strongly
affected by changes in $\epsilon$, $\Omega$, and $\sigma_\omega$. 
Increasing $\epsilon$, decreasing $\sigma_\epsilon$, or decreasing
$\Omega$ by $20\%$ increases $e_{max}$ by $\geq 30\%$. 

\begin{figure}
\plotone{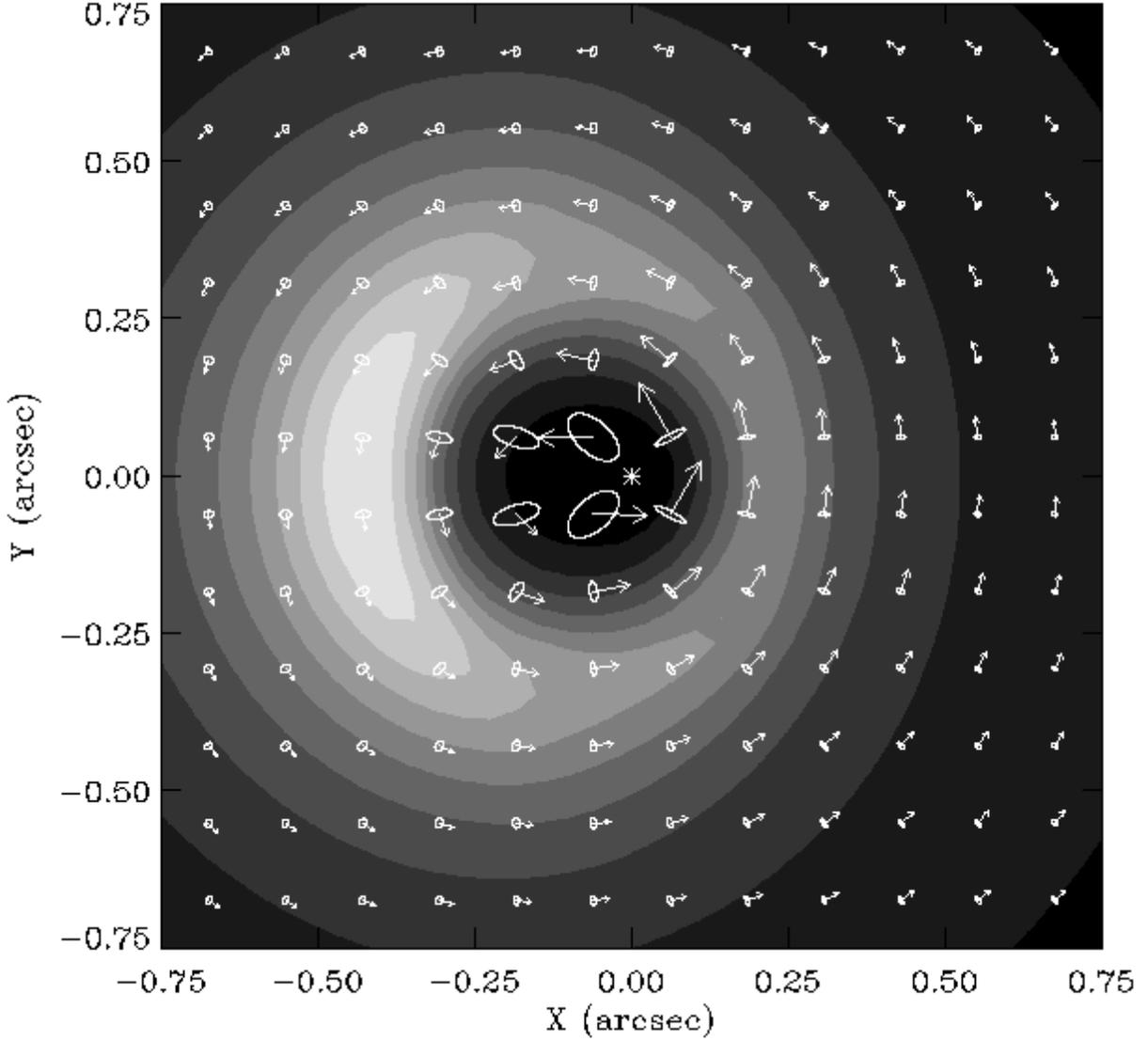}
\caption{\footnotesize
Disk-only mean velocity vectors (arrows) and velocity ellipsoids
(ellipses) plotted over the surface density (contours) for Model 14 in
Table \ref{t.results2}, which has a Gaussian $F(e)$.  Density contours are
at 0.1, 0.2, ..., 1.0 of the maximum density.  An asterisk marks the
location of the BH.  Velocity ellipsoids in our disks are elongated in the
radial direction, as expected from epicycle theory; most ellipsoids have a
vertex deviations of $< 10\arcdeg$, and the maximum deviation is
$\sim 30\arcdeg$.}
\label{f.vellipse14out}
\end{figure}

\begin{figure}
\plotone{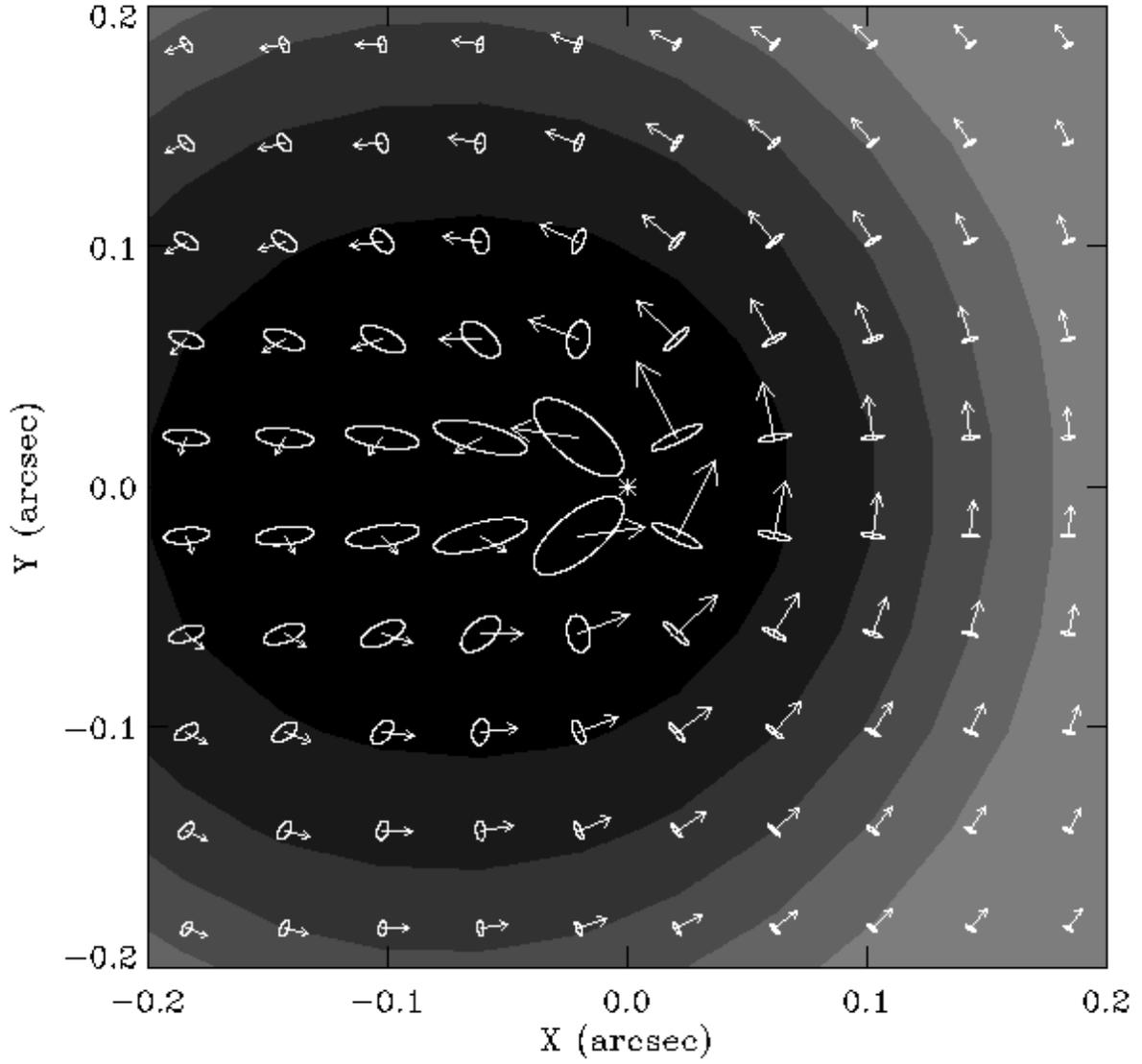}
\caption{\footnotesize
Same as in Figure \ref{f.vellipse14out} for the inner $0\farcs2$.  
Velocities are scaled to 1/5 of their values in Figure 
\ref{f.vellipse14out}.}
\label{f.vellipse14in}
\end{figure}

\begin{figure}
\plotone{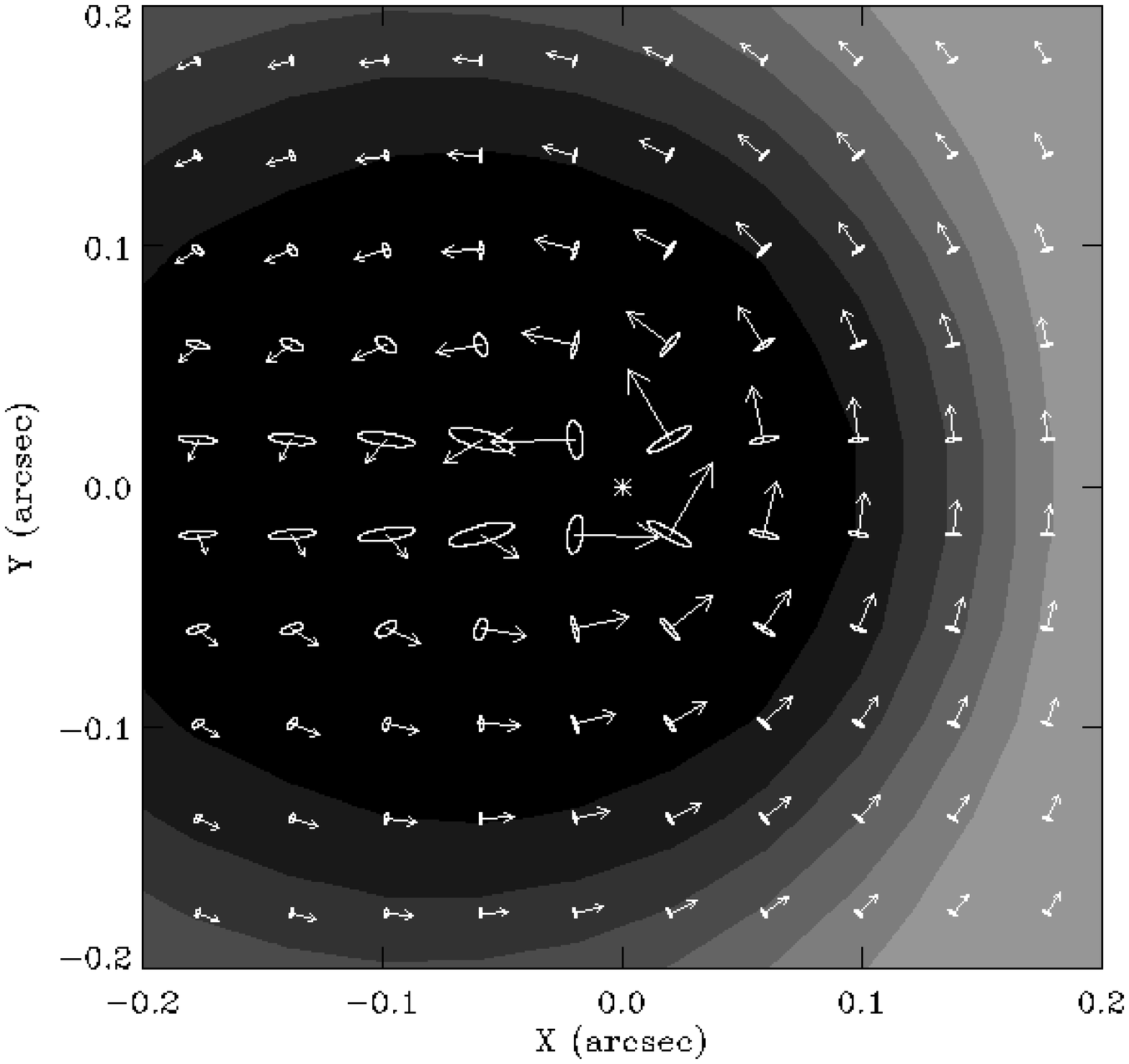}
\caption{\footnotesize
Similar to Figure \ref{f.vellipse14in}, but for the disk in Model 13 in 
Table \ref{t.results2}, which has a Rayleigh $F(e)$.  Comparison with Figure 
\ref{f.vellipse14in} shows that the velocity dispersion 
in a disk with a Gaussian $F(e)$ is larger;  the singularity at $e=0$ in 
the Gaussian distribution causes there to be an extra population of 
circular orbits, in addition to the normal eccentric population about 
$e_0(a)$.
}
\label{f.vellipse13in}
\end{figure}

Figure \ref{f.vellipse14out} shows mean velocity vectors and velocity
ellipsoids for the disk of Model 14, plotted over the density
distribution.  Figure \ref{f.vellipse14in} shows the same inside
$0\farcs2$, with the velocity vectors and ellipsoids scaled by 1/5 of
their values in Figure \ref{f.vellipse14out}.  A figure similar to Figure
\ref{f.vellipse14in} is shown for the disk of Model 13 in Figure
\ref{f.vellipse13in};  Model 13 has a Rayleigh eccentricity distribution. 

\begin{figure}
\plotone{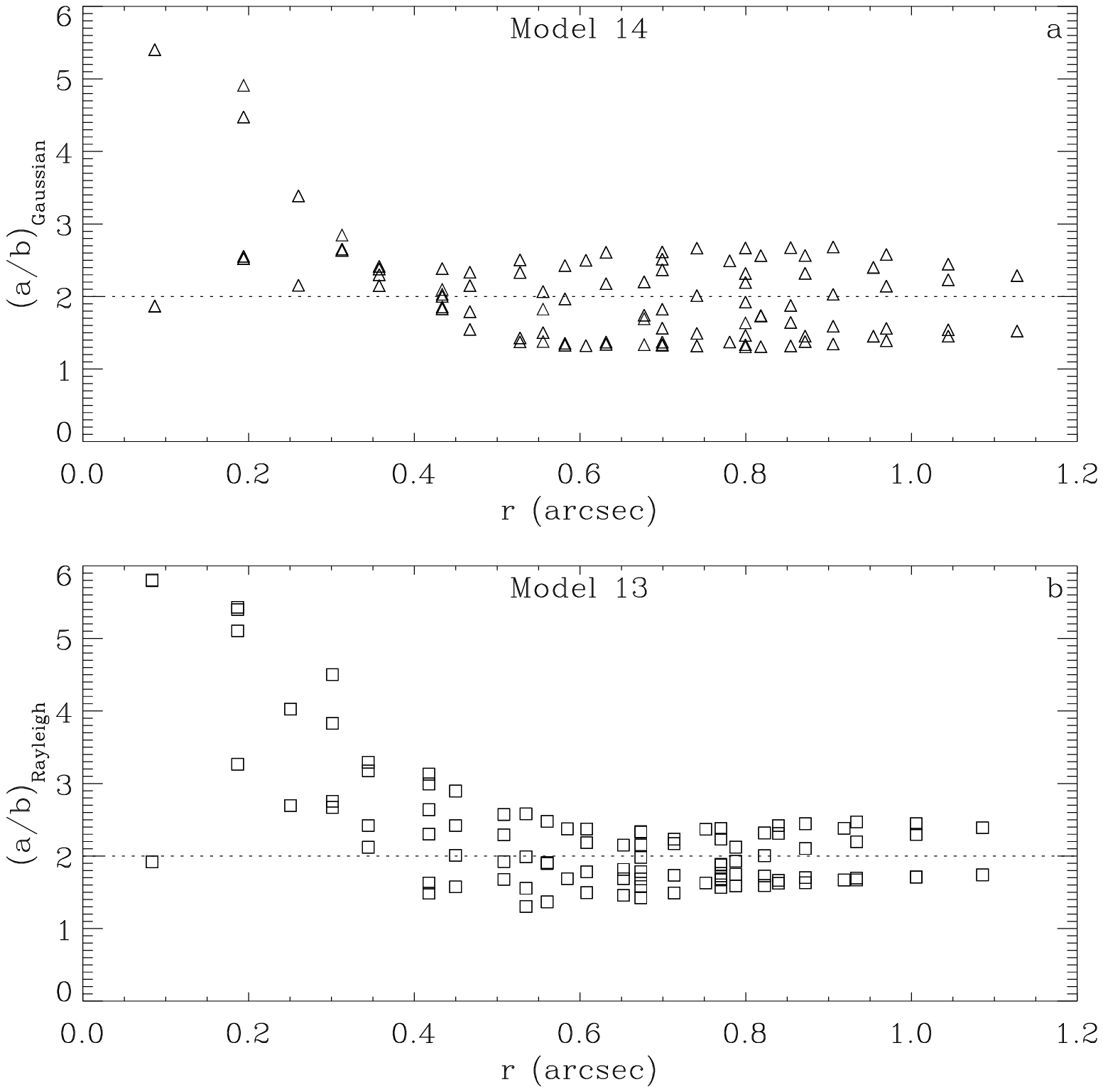}
\caption{\footnotesize
Ratio of major to minor axes for disk-only velocity ellipsoids as a
function of radius from the BH.  (a) For Model 14, which has a Gaussian
$F(e)$.  (b) For Model 13, which has a Rayleigh $F(e)$.  Dotted lines show
an axis ratio of 2, which is expected from epicycle theory for a Keplerian
disk.  The departure from Keplerian behavior occurs near the peak in 
$e_0(a)$ at $a \simeq 0\farcs 3$.
}
\label{f.abratio}
\end{figure}

Velocity ellipsoids are elongated in the radial direction, with
vertex deviations typically less than $10\arcdeg$ and always less than $30 
\arcdeg$.  From epicycle theory, $\sigma_R /
\sigma_T \simeq 2$ for a Keplerian disk, where $\sigma_R$ and $\sigma_T$
are the radial and tangential dispersions, respectively (Binney \&
Tremaine 1987).  Figure \ref{f.abratio}, which plots the ratio of major
to minor axes for velocity ellipsoids as a function of radius, shows that
ellipsoids from Models 14 (panel a) and 13 (panel b) approximately follow
this trend beyond $r \sim 0\farcs4$.  The point of transition away from 
Keplerian behavior occurs near the peak in $e_0(a)$ at $a \simeq 
0\farcs3$.

Comparison of Figures \ref{f.vellipse14in} and \ref{f.vellipse13in} shows
that the disk velocity dispersion is larger in Model 14.  This results
from the singular nature of the Gaussian form of $F(e)$ at $e=0$ (see
Section \ref{s.df}).  The singularity causes there to be two orbit
populations:  first, a normal population of eccentric orbits around
$e_0(a)$, and second, an extra population of circular orbits from the
singularity.  Differences in eccentricity between the two populations
increases the velocity dispersion.  The increase in dispersion is more
prominent at radii where $e_0(a)$ is significantly different than zero. 


\section{Discussion}
\label{sec:CHAP5}

\begin{figure}
\plotone{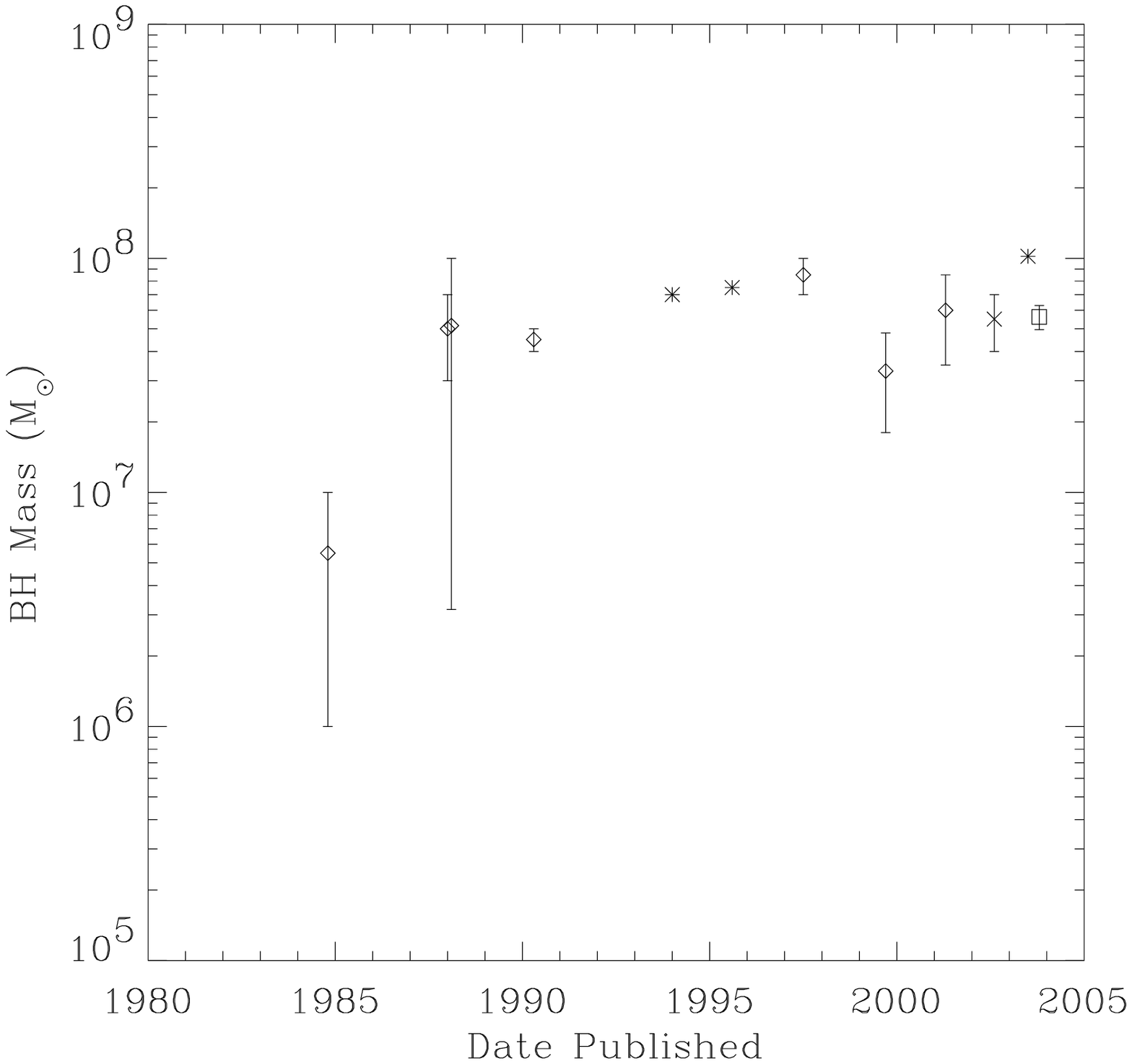}
\caption{\footnotesize
M31 BH mass versus publication date for the values reported in Section
\ref{s.bhresults}.  Diamonds give the median value when a range of BH
masses was published.  Asterisks show BH masses for which no error
estimate or range was provided.  Also plotted is our best-fit value, $5.62
\pm 0.66 \times 10^7 \msun$, denoted by a square, and the value computed
from the $M_{BH}$ - $\sigma$ correlation given in
Tremaine et al. (2002), $5.5 \pm 1.5 \times 10^7 \msun$ (assuming their
value of $160 \pm 8 \kms$ for the dispersion), denoted by a cross.  The 
close agreement between our value and that from the correlation is 
remarkable.
}
\label{f.bhtime}
\end{figure}

We find that the mass of the central BH in M31 is $5.62 \pm 0.66 \times
10^7 \msun$.  To put our result into context, we show BH mass estimates
from various authors (see Section \ref{s.bhresults}) in Figure
\ref{f.bhtime} as a function of publication date.  With the exception of
the Dressler (1984) and KB99 values, all BH mass estimates are consistent
with $M_{BH} \geq 5 \times 10^7 \msun$.  Dressler's value was estimated
from gradients in $M/L$ and $\sigma$, rather than from kinematic
and photometric modeling, and thus is only an order-of-magnitude estimate. 
KB99's value was determined using the displacement of the UV peak relative
to the bulge center, and may be low due to systematic errors in position
measurements (PT03). 

The cross in Figure \ref {f.bhtime} shows the BH mass estimate from the
slope of the $M_{BH}$ - $\sigma$ correlation given in Tremaine et al. 
(2002).  Recall that the authors find $\log(M_{BH}/\msun) = (8.13 \pm
0.06) + (4.02 \pm 0.32) \log(\sigma/200\kms)$, from a sample of $31$
galaxies with reliable BH masses and dispersion measurements.  For M31,
$\sigma = 160 \pm 8 \kms$ (Tremaine et al.  2002), which gives
$M_{BH} = 5.5 \pm 1.5 \times 10^7 \msun$ using the correlation.  The close
agreement between this value and ours is rather remarkable. 

Our BH mass is significantly lower than that found by PT03 in their fit to
bulge-subtracted SIS/CFHT data (KB99) using Monte-Carlo simulations of
eccentric disks built from non-interacting Kepler orbits.  They find
$M_{BH}=10.2 \times 10^7 \msun$ for their non-aligned model, in which the
orientation of the disk is fitted to the data; error bars are not given
with this measurement.  This value is more than $60\%$ larger than the
upper limit for our measurement.

Understanding this discrepancy is difficult, due to fundamental differences
in model approximations.  PT03 ignore disk self-gravity and
precession, but include the three-dimensional structure of the disk;  our
disks include self-gravity and precession, but are assumed to be
thin, and limited to two dimensions.  PT03 suggest that ignoring the disk
self-gravity is likely to cause the BH mass to be $10-20\%$ too large. 
However, it is unknown how including both self-gravity and precession
together would further affect their mass value, especially if the
precession rate is as large as we find in our models ($36.5 \pm 4.2
\kmspc$).  Similarly, it is difficult to ascertain how including a
vertical structure to our disks would affect $M_{BH}$, $\Omega$, and
possibly other parameters. 

Results from PT03 suggest that some vertical structure is needed to match
the photometry.  Their non-aligned model reproduces the well-defined
double structure of M31's nucleus, whereas our two-dimensional simulations
produce crescent-shaped P1 structures.  The dip in surface brightness
between P1 and P2 is reproduced in their models, even with the disk at
$i=54\arcdeg$;  our models possess overly-strong central surface
brightness minima, unless a strong bulge cusp is included in the model. 
PT03 also give dynamical reasons for having vertical structure in the
disk.  Using results from studies of disk heating by two-body
relaxation in protoplanetary disks (Ohtsuki, Stewart, \& Ida 2002), they
show that their non-aligned disk must have a vertical-to-radial axis ratio
of $\sim 0.25$ in the radius range $0.5-1\arcsec$.  This being said, PT03
point out that the effect on the kinematics is less dramatic than for the
photometry.  Since weighting of the data is dominated by one and
two-dimensional kinematics in our best fits (Table \ref{t.results2}), our
value for the BH mass should suffer only minor modulation with the
addition of the third dimension.  Vertical dispersion would increase the
overall line of sight velocity dispersion, while minimally affecting the
rotation curve;  this would improve the fit in models like Model 2 (
Figure \ref{f.1dkinbest1}), which has a small dispersion spike. 

\begin{figure}
\plotone{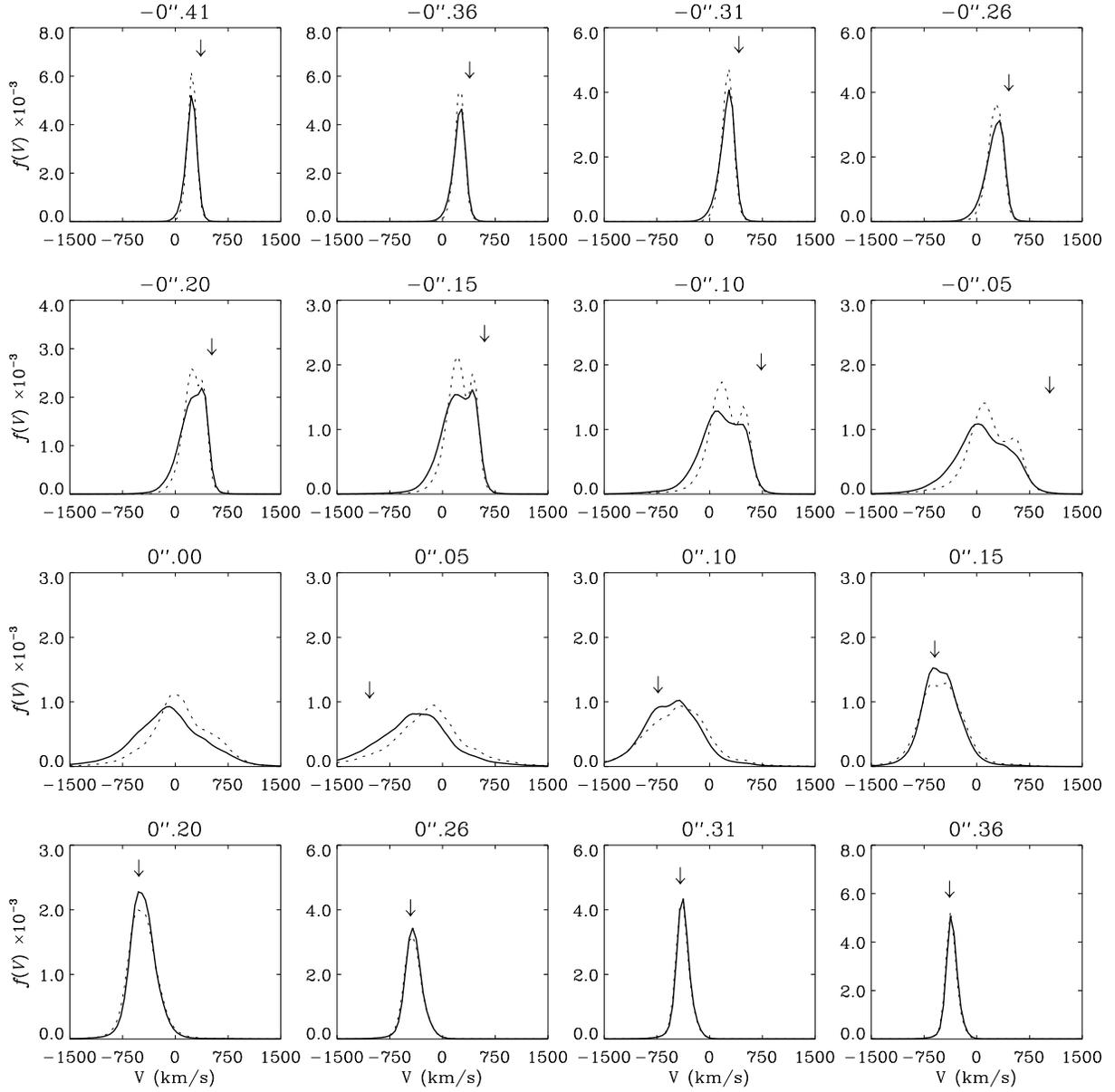}
\caption{\footnotesize
Disk-only LOSVDs near the UV peak for Model 14 (solid lines) and Model 13
(dotted lines) from Table \ref{t.results2}, for the $0\farcs 1$ wide STIS 
slit along {\it PA}$=39\arcdeg$.  The distance from the UV peak along
the slit is given above each panel.  Arrows mark the circular speed at the
tangent point.  These are to be compared with LOSVDs from unpublished 
STIS observations (Bender et al. 2003) presented in PT03 (their Figure 
15).  Both model and data show a small bump at supracircular velocities
in the $0\farcs 10$ and $0\farcs 15$ Panels;  these result from the 
density and eccentricity structure of the disk.
}
\label{f.losvds1413}
\end{figure}

PT03 present bulge-subtracted LOSVDs from unpublished STIS observations of
M31's nucleus (Bender et al.  2003) at a few locations within $\pm 0\farcs
15$ of the UV peak, along with model LOSVDs extending another $\pm 0\farcs
25$, in their Figure 15.  We show corresponding LOSVDs for Models 14 and
13 as solid and dotted lines in Figure \ref{f.losvds1413}, respectively. 
LOSVDs from Bender et al. possess multiple maxima, some of which may be
real features.  Of particular interest are their LOSVDs at $0\farcs 10$
and $0\farcs 15$, both of which show a small bump near $v = 750
\kms$;  our LOSVDs also show these bumps, which occur at supracircular
velocities, as indicated by the arrows in our Figure \ref{f.losvds1413}. 
Supracircular peaks occur when the tangent point falls near the
pericenters of orbits with substantial eccentricity.  Such features arise 
from the density and eccentricity structure
of the disk, and can be used as sensitive discriminants of disk structure 
in M31 (Salow \& Statler 2001).

Model LOSVDs from PT03 do not have multiple maxima; instead, they find 
asymmetric LOSVDs with strong wings toward prograde velocities.  However, 
their LOSVDs between $0\farcs10$ and $0\farcs20$ have a shoulder that 
appears to move inward toward $v = 0 \kms$ in the same manner as the bump 
at supracircular velocities does in our LOSVDs.  This may be a signature 
of the tangent point traversing the pericenters of eccentric orbits, with 
the gap between the maxima somehow filled in by the density structure 
of the disk.

\begin{figure}
\plotone{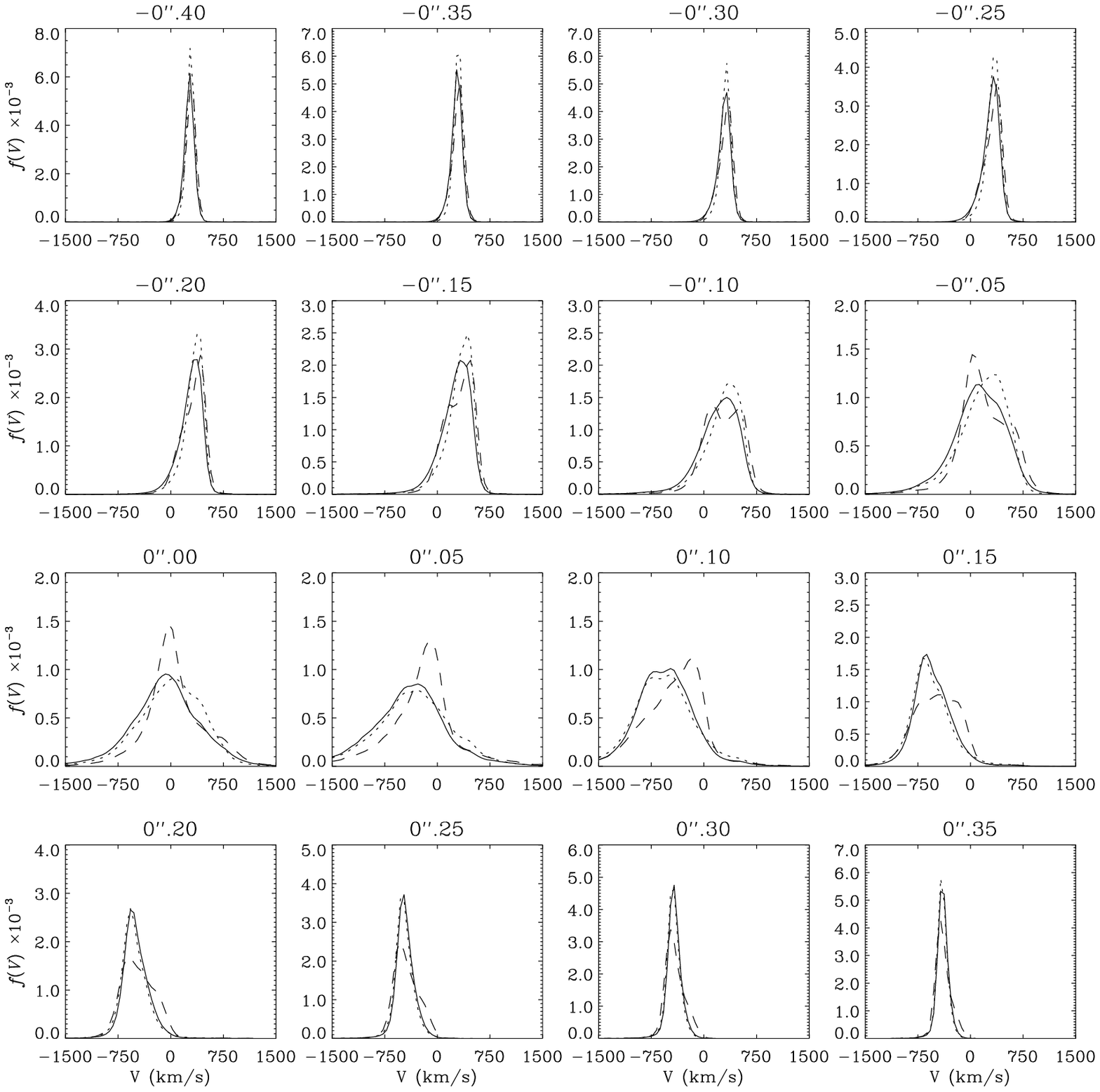}
\caption{\footnotesize
Disk-only LOSVDs for Model 14 (solid lines), Model 12 (dotted lines), and
Model 10 (dashed lines) from Table \ref{t.results2}, for the $0\farcs 1$
wide STIS slit along the kinematic axis ({\it PA}$=56.4\arcdeg$).  The 
distance
from the UV peak along the slit is given above each panel.  $M_{BH}$,
$M_d$, and $\Omega$ all increase from Model 10 to Model 14 to Model 12; 
Model 10 is at $i=68\arcdeg$, compared to $i=52.5\arcdeg$ for the other
two.  These are shown as predictions for upcoming STIS observations at
$S/N \sim 120$.  The three models can be distinguished between $-0\farcs
10$ and $0\farcs 05$, so with high $S/N$ it should be possible to
differentiate between models using LOSVDs. 
}
\label{f.losvdsprop}
\end{figure}

We show disk-only LOSVDs at the resolution of STIS along the kinematic
axis ({\it PA}$_K = 56.4\arcdeg$) in Figure \ref{f.losvdsprop} for Models 
14, 12,
and 10 from Table \ref{t.results2}, which have BH masses of $5.55 \times
10^7 \msun$, $6.84 \times 10^7 \msun$, and $4.67 \times 10^7 \msun$,
respectively.  These are shown as predictions for upcoming STIS
observations, which should yield $S/N \sim 120$ in the $4500-5500 \AA$
region, and allow detailed features in the LOSVD to be seen (Cycle 12
ID-9859, E. Emsellem, PI).  $M_{BH}$, $M_d$, and $\Omega$ all increase
from Model 10 to Model 14 to Model 12.  The most significant differences
in LOSVDs are seen between Model 10 and the other two, mostly due to the
$\sim 16\arcdeg$ inclination difference.  LOSVDs for Models 14 and 12 are
similar throughout much of the near-UV peak region, though there are
measurable differences in the number of maxima and their strength and
location, especially between $-0\farcs10$ and $0\farcs05$.  Thus, with
high S/N, it should be possible to differentiate between models using
LOSVDs. 

The Bender et al. (2003) velocity dispersion profiles presented in PT03
(their Figure 12) deserve special mention.  The dispersion spike in their
bulge-subtracted STIS profile is offset from the UV peak by only $0\farcs
08$, compared to the $\sim 0\farcs2$ offset found by B01 in both STIS
(their Figure 11) and slit-averaged OASIS (their Figure 8) dispersion
profiles.  The large offset cannot be reproduced by our models.  But the
$\sim 0\farcs1$ discrepancy suggests that there may be a
problem with the positional registration of the data in either the Bender
et al. (2003) or B01 results.  The Bender et al. results are
bulge-subtracted, unlike in B01, but the addition of a bulge component
should move the dispersion spike slightly closer to the UV peak, not away
from it, if the BH is centered in the bulge.  Bender et al.'s profile is
favored by our models, which typically have the dispersion spike only
slightly offset toward P2;  for example, the dispersion profile for Model
10 (Figure \ref{f.1dkinbest2}) has a dispersion spike offset of $\sim
0.03$ from the UV peak. 

If the dispersion spike really is offset from the UV peak by $\sim
0\farcs2$ toward P2, then either our models are essentially correct but
missing a key ingredient, or the basic assumptions of the model
incorrectly describe M31's nucleus.  This second possibility is unlikely,
given that our models are able to reproduce many of the key features in
both the kinematic and photometric data; the same holds true for models
presented in T95, KB99, B01, Salow \& Statler (2001), SS02, Jacobs \& 
Sellwood (2001), and PT03.  As for the first
possibility, one suggestion is the addition of retrograde orbits.  Our
models include only prograde orbits.  SS02, who build
their model using an orbit library with both prograde and retrograde
orbits, find that the fit near P2 is improved greatly with the addition of
retrograde orbits comprising only $3.4\%$ of the total disk mass. 
However, the dispersion spike in their model is located very close to the
UV peak, even with retrograde orbits (see their Figure 4).  As a rough
test, we added velocity moments from retrograde orbits comprising $0.05
M_d$ to a few models before convolution with the PSF.  A set of retrograde
periodic orbits was found numerically in the same way as was done for the
prograde orbits, and the moments were found using the DF, as in Section
\ref{s.construction}.  We similarly found that the dispersion spike did not
significantly move.  It should be noted that these tests were performed
using the same $F(a)$ and other disk parameters, excepting $M_d$, as the
main eccentric disk itself, which may not realistically describe the
hypothetical retrograde population.

Using a simple heuristic velocity model, B01 find that the best overall
fit to FOC, STIS and slit-averaged OASIS kinematics
requires a high velocity component in the central $0\farcs3$ aligned with
the kinematic axis.  This component is added to reproduce the abrupt jump
in the FOC rotation curve near $x=0\farcs 1$ (see Figure
\ref{f.1dkinbest1}).  B01 argue that, with the addition of this component,
the dispersion spike is then just the result of the effect of velocity
broadening.  The high velocity component may be related to retrograde
orbits.  The question of whether or not retrograde orbits can affect the
dispersion spike clearly needs further investigation. 

We find that disks with $i = 52.5\arcdeg$ provide the best match to the
two-dimensional kinematics and photometry.  This inclination is consistent
with that found by deprojecting the nucleus, assuming a thin disk (B01,
SS02, P02), and with PT03's fit for their non-aligned model.  Thus, the
nuclear disk is most likely not aligned with the large-scale disk of M31,
which is at $i=77\arcdeg$ with the line-of-nodes at {\it PA}$_n=38\arcdeg$
(T95).  M31's bulge is thought to be triaxial, and possibly aligned with
the large-scale disk (Lindblad 1956, Stark 1977, Stark \& Binney 1994,
Berman 2001, Berman \& Loinard 2002).  The nuclear disk should then be
subject to dynamical friction from the bulge, which acts to damp the
inclination difference on the precession time ($\sim 10^7$ yr; PT03 and
references therein).  At that timescale, the inclination difference should
have been damped out long ago, since absorption-index radial profiles
suggest that the age of the nuclear disk is roughly $1/3$ that of the
bulge (Sil'chenko et al. 1998), which is on order of a Hubble time. 
However, the recent photometric decomposition by P02 suggests that the
inner bulge may be spherical, rather than triaxial.  P02 finds that the
overall bulge is fit best by two components;  an inner, nearly spherical
component (axis ratio $q = 0.97 \pm 0.02$) described by a S\'ersic
(1968) light profile with effective radius $3\farcs31$ and exponent
$n=0.83$, and a large-scale, more elliptical component ($q = 0.81 \pm
0.01$) described by a Nuker law (Lauer et al. 1995) of break radius
$66\farcs 48$ and an asymptotic inner power law slope $\gamma=0.17$.  The
mass of the spherical component is $M_s = 2.8 \times 10^7 \msun$, roughly
half of the mass of the BH.  If this is correct, then the bulge potential
is spherical around the nucleus, which might allow the non-aligned
orientation to survive, even if the outer bulge is triaxial.  Another
possible way to avoid dynamical friction from the bulge is to have an
axisymmetric bulge which is aligned with the nuclear disk (PT03).  It is
intriguing that Ruiz (1976) finds that an axisymmetric bulge is
consistent with kinematic and photometric data if $i=55\arcdeg$. 

All of the models presented in this paper have a backbone orbit sequence,
$e_0(a)$, similar to that shown for Model 14 (the solid line in Figure
\ref{f.densorbeofa}{c}), in which there is no tendency for the orbits to
switch their apoapses to the P2 side of the disk following the large
negative eccentricity gradient.  This is contrary to the the findings of
S99 and Salow \& Statler (2001).  B01 and SS02 also find
$e(a)$s similar to that for Model 14, though with a larger maximum
eccentricity ($e_{max} \geq 0.6$).  Many of our models do, however, show
such a switch at low semimajor axis, before the maximum in $e_0(a)$ (see the
dotted line in Figure \ref{f.densorbeofa}{c}), as in Salow \& Statler
(2001).  Tremaine (2001) finds such a switch on both sides of a
maximum in $e(r)$, where $r$ is the radius from the central massive object,
for certain slow $p$-modes in nearly Keplerian disks with softened
gravity, using the WKB approximation (see his Figures 6 and 9); 
self-gravitating disks with significant velocity dispersion support
prograde ``pressure'' modes, corresponding to Tremaine's $p$-modes (B01). 
Models like ours can, in principle, possess $e_0(a)$s with an
eccentricity sign switch after maximum, or on both sides of maximum, but
only if $\Omega$ is decreased by about $40\%$ from the values we find in
our fits to M31.  For example, Figure \ref{f.extendeofa}{a} in Appendix
\ref{App_2} shows $e_0(a)$ for an arbitrary model with $\epsilon=0.2$ and
$\Omega \simeq 12 \kmspc$. 

The small value for $e_{max}$ in our models ($0.21 \pm 0.05$) is a
consequence of the large precession rate ($36.5 \pm 4.2 \kmspc$). 
Lagrange's planetary equations for the secular evolution orbital elements
undergoing an external perturbation have $\Omega \simeq (1/n a^2 e)
(\partial R / \partial e)$, where $n$ is the mean motion and $R$ is the
Disturbing Function, or perturbing potential (Murray \& Dermott 1999); 
thus, $e \sim 1/\Omega$.  Models in B01 and SS02 have larger values for
$e_{max}$ ($\sim 0.7$) and lower precession rates ($3 \kmspc$ and $16
\kmspc$, respectively), in agreement with the rough trend $e_{max} \sim
1/\Omega$.  It is interesting to note that the precession rate measured
using the Tremaine \& Weinberg (1984) method gives $\Omega = 34 \pm 8
\kmspc$ and $\Omega = 20 \pm 12 \kmspc$ for a Nuker and S\'ersic
fit to the bulge (Sambhus \& Sridhar 2000), respectively, both of which
are consistent with our value for $\Omega$. 

The question of how the eccentric disk formed in M31 is still an open
question.  Currently, two formation scenarios are favored:  first, that an
initially axisymmetric disk becomes lopsided due to an external
perturbation (T95, B01, P02, PT03), or by a dynamical instability (Touma
2002, SS02);  second, that the disk is formed when an infalling star
cluster is tidally stripped by the central BH (Bekki 2000, Quillen \&
Hubbard 2003).  The external perturbation may come from a globular cluster
or giant molecular cloud passing by the disk (B01, P02), or by the
influence of dynamical friction from the bulge (T95, PT03), both of which
may excite the mean eccentricity of the disk.  Using a softened analogue
of the Laplace-Lagrange secular theory for interacting planar Keplerian
rings, Touma (2002) showed that a small fraction of counter-rotating stars
is sufficient to cause a pre-existing disk to develop a linear $m=1$
instability.  The retrograde orbits may have originated from a tidally
disrupted stellar cluster on a retrograde orbit (SS02);  Tremaine et al.
(1975) have shown that dynamical friction can cause globular clusters to
spiral into the nucleus and be tidally disrupted. 

Bekki (2000) performed N-body simulations in which a globular cluster is
disrupted by a massive BH, and found that a long-lived eccentric disk can be
produced.  The progenitor cluster, however, was only about one-tenth as
massive as the disk in M31.  From simple tidal disruption arguments in a
single disruption-event scenario, Quillen \& Hubbard (2003) show that many
Galactic globular clusters satisfy core radius and density requirements
necessary for the formation of an eccentric disk near the BH, supporting
Bekki's simulations.  Normal globular clusters are not massive enough,
however, to be plausible progenitors, and their colors are unlike 
those in the nucleus.  Instead, Quillen \& Hubbard suggest that a dense
bulge core or nuclear star cluster might be the progenitor, since both can
be massive and compact enough to satisfy requirements for M31.  They
point out that if merging galaxy bulges can form eccentric disks, then
such disks would be a natural consequence of hierarchical galaxy
formation. 

An interesting connection may exist between the disruption-event scenario
and the spherical inner bulge found by P02.  Milosavljevic \& Merritt
(2001) perform N-body simulations of merging stellar systems with black
holes, and find that the inward-spiraling binary black holes scatter stars
from the center via gravitational slingshot.  P02 suggests that these
ejected stars could form a spherical distribution after redistribution in
phase space.  If this is true, then the presence of the spherical bulge
component would lend support to the idea that the disk formed during a 
merger of galaxies with central BHs. 


\section{Conclusion}
\label{sec:CHAP6}

Our models of eccentric stellar disks around central black holes
incorporate self-gravity, finite velocity dispersion, and gravity-induced
precession, in a self-consistent way.  We have used these models to
perform the first detailed fit to the nucleus of M31 which includes both
one and two-dimensional kinematics and photometry;  the data set includes
FOC, STIS, and SIS one-dimensional kinematics, OASIS two-dimensional
kinematics, and one and two-dimensional WFPC2 photometry.  The primary
result of this modeling effort is an accurate measurement of the mass of
the central black hole in M31.  We find that $M_{BH} = 5.62 \pm 0.66
\times 10^7 \msun$.  This value is consistent with the $M_{BH} - \sigma$
correlation (Tremaine et al. 2002), which gives a value of $5.5 \pm 1.5
\times 10^7 \msun$.

We find eccentric disks with large precession rates ($\Omega = 36.5 \pm
4.2 \kmspc$) and small maximum eccentricities ($e_{max} = 0.21 \pm 0.05$) 
for the backbone orbit sequence, $e_0(a)$.  The backbone orbits possess a
characteristic non-monotonic distribution with a steep negative
eccentricity gradient ($de/da < 0$) through the densest part of the disk,
which gives rise to distinctive multi-modal LOSVDs for lines of sight near
the central BH.  Such features may be used to further constrain model
parameters when LOSVDs from upcoming high S/N STIS observations of M31's
nucleus become available. 

Although our models provide an accurate estimate for the BH mass, there is
room for improvement.  We have assumed that the disk in M31 is thin. The
disk may have non-negligible vertical structure, however, which could
slightly alter our BH estimate.  Vertical velocity dispersion can be added
to our models by including dispersions in inclination, $i$, and in the
longitude of the ascending node, $\Omega_n$, in our prescription for
populating quasi-periodic orbits about the backbone orbit sequence.  We 
would then have a DF which includes all five integrals of motion in the
three-dimensional Kepler problem; that is, $f(a,e,\omega,i,\Omega_n)$; 
PT03 demonstrate how the third dimension can be included in this way. 
Further improvement may require that new bulge models be considered, to
better fit the behavior of the models within the central $0 \farcs 4$ of
the nucleus.  More flexible versions of $F(a)$ and a population of
retrograde orbits can also help in this regard.  Retrograde orbits should
be included self-consistently with their own DF and parameters, to
determine if the location of the dispersion spike can be adjusted by their
presence. 

The eccentric disk picture for the nucleus of M31 is clearly the correct
one, given the success that our models, along with those of other
investigators, have at reproducing most of the asymmetric features in the
kinematic and photometric data.  A complete description of the available
data should be possible with models like ours, which already include most
of the key physical ingredients, once they are extended as mentioned
above.  Such models should be flexible enough to probe the connection
between the BH and nuclear stars in greater detail, and may yield new
clues about the formation of eccentric disks around BHs.  With the
knowledge gained from detailed study of these models, other systems
exhibiting features like those seen in M31, which do not have resolved
Keplerian-dominated regions ($r_k$), can be investigated with confidence,
as long as the sphere of influence of the BH ($r_h$) is resolved. 

Galaxies with central properties similar to those in M31 are already known
to exist.  NGC 4486B, a low-luminosity E1 companion of M87, is already
known to possess two brightness peaks (Lauer et al. 1996).  This galaxy
has no distinct nucleus, however, and its P1-P2 separation is $\sim 10$ -
$13$ pc ($\sim 0\farcs 15$ at $D=16$ Mpc), which is about six times that
in M31;  also, its P1 and P2 have nearly the same brightess, unlike in
M31.  Models like ours may be applicable, with some modification to the
density structure to account for lack of a distinct nucleus.  Dynamical
modeling of spectroscopic data from SIS/CFHT suggests that NGC 4486B
harbors a BH of mass $\sim 6 \times 10^8 \msun$ (Kormendy et al. 1997). 
Kormendy et al. find $\sigma \simeq 130 \kms$, which, along with the
aforementioned BH mass, implies that $r_h \simeq 2\arcsec$ (probably as an
upper limit), which can be resolved by STIS/HST.  Further examples are
shown by Lauer et al. (2002), who recently discovered six early-type
galaxies with surface brightness profiles that decrease inward near their
centers, reminiscent of the central dip in surface brightness found in
M31.  These galaxies harbor torus-like brightness distributions, rather
than double nuclei. Such structures can possibly be fit by a thin disk
with low $\epsilon / \Omega$, which is only slightly asymmetric.  The
sharpest structure observed in the sample is in the S0 galaxy NGC 3706,
which has a bright stellar torus of radius $r \sim 20$ pc ($0\farcs 12$ at
$D = 35$ Mpc).  Using the velocity dispersion observations from Carollo \&
Danziger (1994), the $M_{BH}$ - $\sigma$ correlation gives $M_{BH} \simeq
6 \times 10^8 \msun$ for this galaxy.  With this mass, $r_h \simeq 0\farcs
15$.  Modeling this galaxy will require observations with higher
resolution than is currently available, but which may be available in the
near future. 

\acknowledgements
This work was supported by NSF CAREER grant AST-9703036 and Space 
Telescope Science Institute grant HST-GO-08589.02-A.  We thank Eric 
Emsellem for providing useful information and advice and Tod Lauer, John 
Kormendy, Joe Shields, and Brian McNamara for helpful comments.


\appendix
\section{\sc Extending the Orbit Sequence}
\label{App_2}

Beyond the 2:1 resonance it becomes difficult to find nearly-elliptical
periodic orbits using the method described in Section \ref{s.construction}.
Periodic orbits beyond this resonance belong to various resonant
families.  Since the 2:1 resonance falls within $R_d$ for models with
larger values of $\Omega$, we must approximate the backbone structure of
the disk in those cases.  When truncation occurs, we simply assume that
the sequence of nearly-Keplerian orbits continues out to $R_d$.  We do
this in a way that mimics the behavior of $e_0(a)$ for those models whose
orbit sequence does not truncate, using the decaying oscillatory function

\begin{equation}
E(a) = \exp \left( - A a^2 \right) \cos \left( B a + C \right).
\label{e.extend}
\end{equation}

\begin{figure}[tb]
\plotone{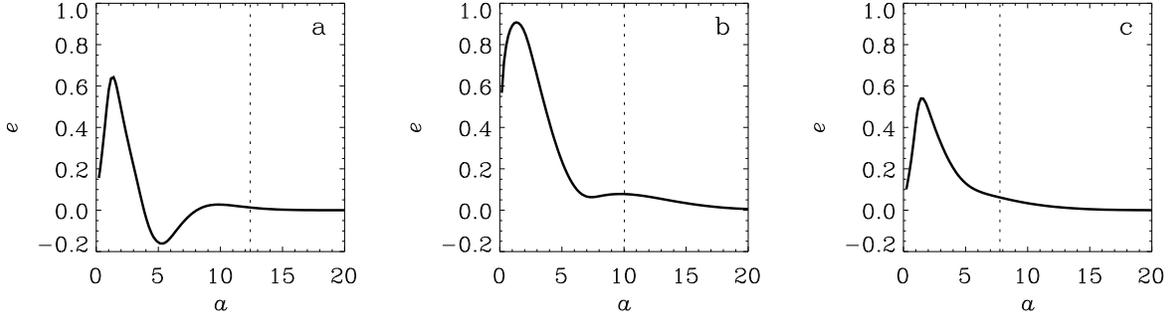}
\caption{\footnotesize
Three examples of extended $e_0(a)$ functions, using $E(a)$ in Equation
\ref{e.extend}.  The dotted lines show the point where the orbit sequence
truncates, near the 2:1 resonance.  The three models shown are not from
the grid of 24 best-fit models;  they are arbitrary models showing three
different types of behavior typically found for $e_0(a)$ in the M31-like
region of parameter space.}
\label{f.extendeofa}
\end{figure}

\noindent
We fit this function to the last $10$ orbits in $e_0(a)$, just before the 
cutoff. Figure \ref{f.extendeofa} shows three examples of extended 
$e_0(a)$ functions for arbitrary models.  The disk structure and dynamics 
was 
found to be insensitive to the choice of the extending function for a wide
range of decaying functions.

\end{document}